\documentclass[journal]{IEEEtran}
\pdfoutput=1 
\usepackage{subfig}
\usepackage{bm}
\usepackage{pgfplots}
    \usetikzlibrary{
        arrows.meta,
        spy,
    }
\usepackage{filecontents}
\usepackage[scientific-notation=true]{siunitx}
\usepackage{amsmath}
\usepackage[T1]{fontenc}
\usepackage{graphicx,array,amsthm,stfloats,upref,amsfonts,amssymb}
\usepackage{color,soul}
\graphicspath{{figures/}}

\newcommand{\spaceto}[2]{{\ooalign{$\phantom{#1}$\cr\hidewidth$#2$\hidewidth}}}
\usepackage{cite}
\usepackage{mathtools}
\DeclarePairedDelimiter\floor{\lfloor}{\rfloor}

\usepgfplotslibrary{polar}

\pgfplotsset{compat=1.13}

\begin{document}
\title{Modeling Aspects of Planar Multi-Mode Antennas for Direction-of-Arrival Estimation}

% \author{Michael~Shell,~\IEEEmembership{Member,~IEEE,}
%         John~Doe,~\IEEEmembership{Fellow,~OSA,}
%         and~Jane~Doe,~\IEEEmembership{Life~Fellow,~IEEE}% <-this % stops a space
% \thanks{M. Shell was with the Department
% of Electrical and Computer Engineering, Georgia Institute of Technology, Atlanta,
% GA, 30332 USA e-mail: (see http://www.michaelshell.org/contact.html).}% <-this % stops a space
% \thanks{J. Doe and J. Doe are with Anonymous University.}% <-this % stops a space
% \thanks{Manuscript received April 19, 2005; revised August 26, 2015.}}
% % \supertitle{Submission Template for IET Research Journal Papers}

\author{Sami~Alkubti~Almasri,
        Robert~P\"ohlmann,
        Niklas~Doose,
        Peter~A.~Hoeher,
        and~Armin~Dammann
\thanks{S. A. Alamsri, N. Doose, and P. A. Hoeher are with the Institute of Electrical Engineering and Information Theory, Kiel University, Kiel 24143, Germany. (e-mail: saaa, nd, ph@tf.uni-kiel.de)}
\thanks{R. P\"ohlmann and A. Dammann are with the Institute of Communications and Navigation, German Aerospace Center (DLR), Wessling 82234, Germany. (e-mail: Robert.Poehlmann, Armin.Dammann@dlr.de)}}

% \address{\add{1}{Institute of Electrical Engineering and Information Theory, Kiel University, Kiel, Germany}
% \add{2}{Institute of Communications and Navigation, German Aerospace Center, Oberpfaffenhofen-Wessling, Germany}
% \email{saaa@tf.uni-kiel.de}}
\maketitle
\begin{abstract}
Multi-mode antennas are an alternative to classical antenna arrays, and hence a promising emerging sensor technology for a vast variety of applications in the areas of array signal processing and digital communications.
An unsolved problem is to describe the radiation pattern of multi-mode antennas in closed analytic form based on calibration measurements or on electromagnetic field (EMF) simulation data. 
As a solution, we investigate two modeling methods: One is based on the array interpolation technique (AIT), the other one on wavefield modeling (WM). 
Both methods are able to accurately interpolate quantized EMF data of a given multi-mode antenna, in our case a planar four-port antenna developed for the 6--8.5 GHz range.
Since the modeling methods inherently depend on parameter sets, we investigate the influence of the parameter choice on the accuracy of both models.
Furthermore, we evaluate the impact of modeling errors for coherent maximum-likelihood direction-of-arrival (DoA) estimation given different model parameters. 
Numerical results are presented for a single polarization component.
Simulations reveal that the estimation bias introduced by model errors is subject to the chosen model parameters.
Finally, we provide optimized sets of AIT and WM parameters for the multi-mode antenna under investigation. 
With these parameter sets, EMF data samples can be reproduced in interpolated form with high angular resolution.
\end{abstract}

\begin{IEEEkeywords}
characteristic modes, array interpolation technique, wavefield modeling, transformation error, direction-of-arrival, maximum-likelihood estimation.
\end{IEEEkeywords}

\section{Introduction}\label{sec1}
A multi-mode (MM) antenna is a physical radiator that is capable of exciting several modes separately. 
Each mode is assigned a different radiation pattern, i.e, several radiation patterns can be emitted simultaneously. 
Particularly in wideband antenna designs each port may excite several modes with different weights, but in narrowband designs each mode approximately corresponds to one port.

The theory of characteristic modes (TCM) \cite{Garbacz71, Harrington71, Chen15} is a versatile design and analysis tool that establishes a theoretical framework describing MM antennas, beside other antenna types. 
Based on this concept, the surface current on an electric conductor can be decomposed into a set of orthogonal components, called characteristic modes.
Each of these modes yields a distinct radiation pattern of the electric far-field. 
Compared to traditional antennas, where only the fundamental mode or a mixture of modes are excited, an MM antenna offers properties of an antenna array given a single physical element.
Hence, an $M$-port MM antenna mimics an antenna array with $M$ elements. 

In a special issue on the theory and applications of characteristic modes published recently \cite{Lau16}, a trend of fast-growing interest in the field has been reported. 
Up-to-date publications, see for example \cite{Salih17, Wen17, Dicandia17, Liang17, Kim18, Dicandia18, Li18}, support this trend. 
It was proven in \cite{Svantesson02} for the first time that for properly designed MM antennas the correlation is low enough to yield significant diversity gain when the MM antenna is used for MIMO transmission.
In the remainder of this contribution, emphasis will be on planar MM antennas exploiting a printed circuit board as a radiator.
It has been shown in \cite{martens14, Manteuffel16, Hoeher17} that compared to linear arrays of the same size as a planar MM antenna, less correlation exists between signals radiated from different ports.
This property in conjunction with a small form factor and a robust structure makes this type of MM antenna attractive for various applications.
Several articles on the performance of the MM antenna under investigation have been published regarding communication aspects, including \cite{Hoeher17, HoeMan17, Doose18}. 
In contrast, our work is dedicated to MM antennas regarding positioning aspects.
An application of the theory of characteristic modes for positioning purposes has recently been proposed in\mbox{\cite{Ma18}}. The authors exploit the chassis of an airborne platform to excite characteristic modes and employ these modes as antenna elements.
Contrarily, our focus is on modeling aspects for compact-size planar MM antennas.

Estimating the direction of arrival (DoA) of incoming electromagnetic waves using an antenna array has been a key technology for decades.
Since the beginning of interest in direction finding, DoA has found applications in various fields like radar, sonar, and navigation.
Being part of the array signal processing field, DoA estimation is theoretically and practically well established and documented in literature, but is still an active research area.
Among variously proposed DoA algorithms, the most important ones are maximum-likelihood techniques and subspace techniques \cite{Tuncer09}. 
Maximum-likelihood techniques are optimal in sense of the mean-squared error (MSE) of the estimated DoAs \cite{Wax83, Bohme86, Jaffer88}. 
On the other hand, a drawback is the high computational complexity, because these techniques perform a multi-dimensional search. 
Subspace methods offer improved computational efficiency. 
The most popular one is the multiple signal classification (MUSIC) technique \cite{Schmidt86}. 
The popularity of MUSIC comes from the fact that it provides high resolution based on a one-dimensional search.
More computationally efficient subspace methods that are even search-free are for example ESPRIT (estimation of signal parameters via rotation invariance techniques) \cite{Roy89} and a variation of MUSIC called Root-MUSIC \cite{Friedlander93}.

While ESPRIT and Root-MUSIC offer good performance at low computational cost, they assume an ideal uniform geometry of the antenna array with a well known response.
In practice, it is difficult to obtain a simple uniform geometry with perfect manufacturing of the antenna elements and typical mounting platforms \cite{Hangqing16}.
This affects the response of the array so it becomes different from the desired one.  
Hence, modeling of antenna arrays has attracted a great amount of attention as a solution for the DoA estimation problem with arrays of arbitrary geometry or arrays with hardware imperfections \cite{Belloni07}.
Among the most popular modeling techniques is the array interpolation technique (AIT) \cite{Friedlander93, Friedlander90, Bronez88} and the wavefield modeling (WM) technique \cite{Doron1, Doron2, Doron3}.
The AIT approach aims into linearly transform the response of the real array with arbitrary geometry to the response of a virtual array with a uniform geometry.
On the other hand, the idea of WM is based on modeling the received wavefield of the real array as an orthogonal expansion to describe the response of the real array analytically.

In previous papers by the authors, initial work on modeling of a planar MM antenna by AIT and WM as well as the suitability of the designed models for DoA estimation has been published.
In \cite{Almasri17}, coherent maximum-likelihood DoA estimation has been investigated using the AIT-based model.
The response of the MM antenna has been transformed to the response of a virtual array with heuristically chosen parameters (e.g. geometry of the virtual array, number of antenna elements, position of the array).
In \cite{Poehlmann17WPNC} and \cite{Poehlmann17CAMSAP}, the WM-based model was applied for DoA estimation based on a non-coherent estimator as well as a coherent one.
Concerning the WM-based mathematical model, in these publications a fixed number of coefficients has been assumed.

In this paper, we carry out an in-depth investigation of modeling aspects of MM antennas for DoA estimation purposes.
Towards this goal, we start by briefly reviving both AIT and WM-based models. 
Next, we study the accuracy of the designed models and the influence of key parameters on the model precision.
We show that with a proper set of parameters both models approximate the MM antenna with high accuracy.
Furthermore, we analyze the influence of the investigated parameters on the DoA estimation performance.
For that purpose we apply the maximum-likelihood technique on the designed models, with varying parameters.
Additionally, we discuss pros and cons of both models with respect to the choice of parameters for each model for the DoA estimation task.
Finally, we provide parameter sets that optimize the MM antenna modeling for both AIT and WM approaches.

 \begin{figure}
	\centering
	\includegraphics[width=\columnwidth,height=0.45\textheight,keepaspectratio]{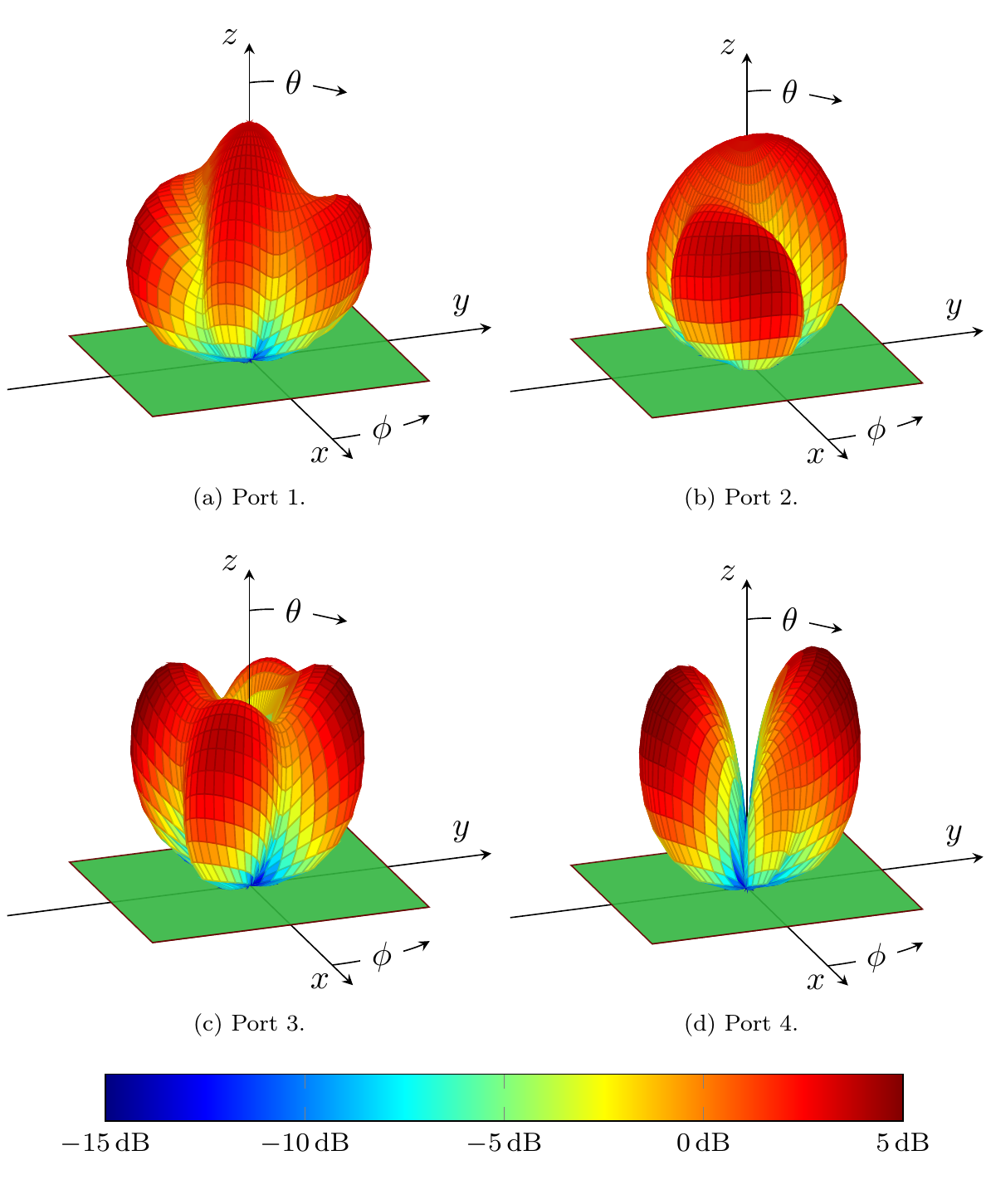}
   \caption{Gain patterns of the investigated MM antenna prototype for right hand circular polarization. The coordinate system under consideration is defined as well.}
   \label{fig1}
 \end{figure} 
 
The following conventions and assumptions hold throughout the paper:  
Numerical results are based on the four-port MM antenna published in \cite{Manteuffel16}, which has been designed for mobile terminals operating in the 6--8.5~GHz band.
Further parameters of this antenna can be found in Table~\ref{table1}.
Additionally, a 484-port antenna has been presented in \cite{Manteuffel16} for application in access points, which assembles 121 four-port elements.
The electric field responses, obtained from the EMF simulations of the MM prototype in \cite{Manteuffel16}, are utilized to calculate the gain patterns.
Fig.~{\ref{fig1}} shows the resulting gain patterns exemplarily for right hand circular polarization (RHCP).
As can be seen in the figure, each of the four ports stimulates a different radiation characteristic, giving $M=4$ distinct modes. 
The planar MM antenna is assumed to radiate only in the upper half of the 3-D space.
For simplicity, we perform the analysis and conduct all simulations in 2-D space.
The feasibility of 3-D DoA estimation with an MM antenna was proven in \cite{Poehlmann17WPNC} and \cite{Poehlmann17CAMSAP}. 
The considered coordinate system, with the MM antenna positioned on the $xy$-plane at the center of the Cartesian coordinate system is depicted in Fig.~\ref{fig1}.
We take the upper half of the $xz$-plane into account, i.e., along the co-elevation angle $\theta\in[-90^\circ,90^\circ]$.
Note that the modeling methods under investigation are applicable to arbitrary polarization components. 
For reasons of conciseness, we show numerical results for RHCP incident signals. %, while assuming that received signals are plain RHCP.
Concerning DoA (and polarization) estimation studies considering diversely polarized incident signals, the interested reader is referred to \cite{Ma18, Costa12} and references therein.

The paper is organized as follows. 
The modeling techniques under investigation are introduced in Section~\ref{sec2}. 
Section~\ref{sec3} is devoted to different practical aspects related to the choice of the approximation model of the MM antenna. 
The influence of modeling parameters on maximum-likelihood DoA estimation is studied in Section~\ref{sec4}. 
The applied modeling techniques and results of the DoA estimation are discussed in Section~\ref{sec5}. 
Finally, conclusions are drawn in Section~\ref{sec6}. 

\begin{center}
\begin{table}[t]
\caption{Physical parameters of the considered MM antenna \cite{Manteuffel16}.}
    \begin{tabular}{ | l | l |}
    \hline
    Radiator dimension & 30 mm $\times$ 30 mm (0.725 $\lambda$)\\
    Center frequency & 7.25 GHz \\
    Number of ports & 4 \\
    Number of coupling elements & 8 \\
    Feed concept & Direct coupling \\
    Feed network & Tri-plate transmission line \\
    \hline
    \end{tabular}\label{table1}
    \end{table}
\end{center}

\section{Multi-Mode Antenna Models}\label{sec2}
Given the fact that an MM antenna is considered to be an antenna array of arbitrary structure, it is important to find a suitable modeling strategy.
A model enables the application of various computationally efficient array signal processing techniques, like for DoA estimation.
For this purpose, the designed model should be able to bear the response of the $m$th port
\begin{equation}
 a_m(\theta) = \sqrt{g_m(\theta)}e^{\mathrm{j}\Phi_m(\theta)}
\end{equation}
for any angle $\theta$, where $g_m(\theta)$ is the antenna gain and $\Phi_m(\theta)$ is the antenna phase response \cite{Balanis07}.
We suggest two different modeling techniques. 
Based on the AIT \cite{Friedlander90}, the first method designs a model that interpolates the radiation characteristics of the MM antenna by means of a virtual array. 
The second method exploits wavefield modeling \cite{Doron1, Doron2, Doron3} in order to design a mathematical model that describes the response of the MM antenna. 
Both techniques are able to inherently interpolate spatially quantized (measured or simulated) EMF data, so the response of the MM antenna can be calculated at any arbitrary angle.
In the following, we adopt the electric field samples from EMF simulations of an MM antenna \cite{Manteuffel16} for constructing both models. 
The available samples are sparsely stored with a step size of $5^{\circ}$ in both azimuth and elevation.

\subsection{AIT-based Model}\label{sec2.1}
The AIT maps the response of an arbitrarily structured array to the response of a virtual uniform array. 
The principle was first introduced by Bronez \cite{Bronez88} to overcome hardware limitations and imperfections, since it is challenging to obtain the desired response of a theoretical uniform array from a real life uniform array. 
His work has been later extended by Friedlander \cite{Friedlander90} and Pesavento \textit{et~al.} \cite{Pesavento02}.  
The model we design is inspired by the AIT concept introduced by Friedlander \cite{Friedlander90}. 
In his work, Friedlander divided the field of view (FoV) into preliminarily defined sectors. 
Next he linearly transformed the response of the arbitrary array to the response of a virtual uniform linear array within each sector. 
Hence, a set of mapping coefficients could be found for each sector. 
Finally, the output of the linear array could be transformed into the output of the arbitrary array by means of the mapping coefficients.

The model we design here, in contrast to conventional AIT, aims to transform the response of a virtual uniform linear array (ULA) to 
the response of the MM antenna. 
Towards this goal we first define the set of $P$ angles, at which the electric field response of the MM antenna prototype is given from EMF simulation data \cite{Manteuffel16} as $\bm{\vartheta}=[\vartheta_1,\vartheta_2,\dots,\vartheta_P]$. 
Next we arrange the complex electric field responses of $M$ ports to a source located at angle $p$ in the vector
\begin{equation}
 \bm{\varepsilon}(\vartheta_p)=[\varepsilon_1(\vartheta_p),\varepsilon_2(\vartheta_p),\dots,\varepsilon_{M}(\vartheta_p)]^\mathrm{T},
\end{equation}
where $\bm{\varepsilon}(\vartheta_p) \in{\mathbb{C}^{M\times 1}}$ and $(\cdot)^\mathrm{T}$ denotes the transpose. Organizing the response 
vectors in a matrix leads to
\begin{equation}
 \bm{E}(\bm{\vartheta})=[\bm{\varepsilon}(\vartheta_1),\bm{\varepsilon}(\vartheta_2),\dots,\bm{\varepsilon}(\vartheta_P)],
\end{equation}
where $\bm{E}(\bm{\vartheta}) \in{\mathbb{C}^{M\times P}}$ contains in each column the complex electric field response of each of the $M$
ports to a signal arriving from angle $\vartheta_p$. The spatial resolution of the given samples should be sufficiently dense, albeit interpolation is performed inherently by AIT or WM processing. 

Now we take a look at the array steering matrix of the virtual ULA. 
The array steering vector $\bm{a}_v(\vartheta_p)\in{\mathbb{C}^{N\times1}}$ of the virtual ULA at angle $\vartheta_p$ can be in general written
as
\begin{equation}
\bm{a}_v(\vartheta_p)=[e^{\mathrm{j}k(x_1\sin{\vartheta_p}+z_1\cos{\vartheta_p})},\dots,e^{\mathrm{j}k(x_N\sin{\vartheta_p}+z_N\cos{\vartheta_p})}]^\mathrm{T},
\end{equation}
where $k$ is the wavenumber and $x_n$  and $z_n$ are the coordinates of the $n$th antenna element in the $xz$-plane. Hence the array 
steering matrix $\bm{A}_v(\bm{\vartheta}) \in \mathbb{C}^{N\times P}$ of the virtual ULA can be expressed as
\begin{equation}
 \bm{A}_v(\bm{\vartheta})=[\bm{a}_v(\vartheta_1),\bm{a}_v(\vartheta_2),\dots,\bm{a}_v(\vartheta_{P})].
\end{equation}
Similar to $\bm{E}(\bm{\vartheta})$, $\bm{A}_v(\bm{\vartheta})$ contains in each column the response of the elements of the virtual ULA to 
a signal arriving from angle $\vartheta_p$. 
The next step is to divide the FoV ($\vartheta\in[-90^\circ,90^\circ]$) into $L$ sectors of equal sizes.
This is necessary because no $\bm{A}_v(\bm{\vartheta})$ can be found for the whole FoV so that the approximation is sufficiently accurate. Evidence is given in Fig.~\ref{fig13}.  
Each sector contains $P_l$
angular samples $\bm{\vartheta}_l=[\vartheta_1^{(l)},\vartheta_2^{(l)},\dots,\vartheta_{P_l}^{(l)}]$. The array steering matrix of the virtual
ULA over the $l$th sector can be expressed as 
\begin{equation}
 \bm{A}_v(\bm{\vartheta}_l)=[\bm{a}_v(\vartheta_1^{(l)}),\bm{a}_v(\vartheta_2^{(l)}),\dots,\bm{a}_v(\vartheta_{P_l}^{(l)})]\subset\bm{A}_v(\bm{\vartheta}), 
\end{equation}
and the electric field responses of the MM antenna over the same sector as 
\begin{equation}
 \bm{E}(\bm{\vartheta}_l)=[\bm{\varepsilon}(\vartheta_1^{(l)}),\bm{\varepsilon}(\vartheta_2^{(l)}),\dots,\bm{\varepsilon}(\vartheta_{P_l}^{(l)})]\subset\bm{E}(\bm{\vartheta}). 
\end{equation}
The problem of finding the mapping coefficients can be solved sector-wise.  We minimize the sum of the quadratic
 errors between the desired response and the interpolated response. This can be described by  
\begin{equation}\label{eq8}
 \bm{\hat{G}}_{l} = \arg\min_{\tilde{\bm{G}_l}}{{\left\| \bm{\tilde{G}}_l^\mathrm{H} \bm{A}_v(\bm{\vartheta}_l)-\bm{E}(\bm{\vartheta}_l) \right\|}_\mathrm{F}^2}, 
\end{equation}
where $\bm{G}_l\in\mathbb{C}^{N\times M}$ is the mapping coefficient matrix corresponding to the sector $l$, $\bm{A}_v(\bm{\vartheta}_l)
\in\mathbb{C}^{N\times P_l}$, and $\bm{E}(\bm{\vartheta}_l)\in\mathbb{C}^{M\times P_l}$.
Throughout this paper, $(\tilde{\cdot})$ denotes a hypothesis, $(\cdot)^\mathrm{H}$ denotes the Hermitian transpose and $||\cdot||_\mathrm{F}$ denotes the Frobenius norm.
It can be seen from \eqref{eq8} that the optimal set of mapping coefficients is obtained by minimizing the squared Frobenius norm with respect to the mapping coefficients, while the virtual ULA parameters (e.g. number of antenna elements, orientation of the array and interelement spacing) are assumed to be given. 
The choice of the latter parameters is discussed in Section~\ref{sec3}.
The solution of problem \eqref{eq8} is known as the least squares solution 
\begin{equation}\label{eq9}
 \bm{\hat{G}}_{l} = \left( \bm{A}_v(\bm{\vartheta}_l)\bm{A}_v^\mathrm{H}(\bm{\vartheta}_l) \right)^{-1}\bm{A}_v(\bm{\vartheta}_l) \bm{E}^\mathrm{H}(\bm{\vartheta}_l),
\end{equation}
where $(\cdot)^{-1}$ denotes the matrix inverse. This solution was found by taking the electric field responses of the MM antenna
and the corresponding virtual ULA array steering vectors at the angles $\vartheta$. After calculating the mapping matrices $\bm{G}_l$ for each sector, 
the array steering matrix of the virtual ULA can be transformed to the interpolated MM antenna response at any set of angles $\bm{\theta}_l$ within sector $l$ 
according to
\begin{equation}
 \bm{A}(\bm{\theta}_l) = \bm{G}_l^\mathrm{H}\bm{A}_v(\bm{\theta}_l).
\end{equation}
For finding the interpolated MM antenna response over the whole FoV ($\theta\in[-90^\circ,90^\circ]$), the responses 
$\bm{A}(\bm{\theta}_l)$ from each sector are concatenated. The model design is performed only once for a given MM 
antenna and certain virtual ULA parameters. The calculation of mapping coefficients for each sector can be performed offline and applied
to map the virtual ULA response to the MM antenna response.   

\subsection{WM-based Model}\label{swm}
The introduction of WM dates back to Doron \textit{et al.} \cite{Doron1}. 
In this paper we only provide a very brief introduction, for the details please refer to \cite{Doron1, Costa10}. 
The general idea is to decompose the antenna response vector
\begin{equation}\label{eq:ms}
\bm{a}(\theta) = \bm{H} \, \bm{\Psi}(\theta) \in \mathbb{C}^{M\times1}
\end{equation}
as the product of a sampling matrix $\bm{H} \in \mathbb{C}^{M \times U}$ and a basis vector $\bm{\Psi}(\theta) \in \mathbb{C}^{U\times1}$, where $\theta\in[-90^\circ,90^\circ]$ is the DoA. The sampling matrix is completely independent of the received wavefield, i.e. the DoA of the signal. The basis vector, on the other hand, is independent of the employed antenna. 
Different choices for the basis functions exist, keeping in mind that they have to be orthonormal on the respective manifold, i.e. $\theta \in [-180^\circ,180^\circ)$. 
A natural choice in this case are the Fourier functions
\begin{equation}\label{eq:Psitheta}
[\bm{\Psi}(\theta)]_u = \frac{1}{\sqrt{2\pi}}e^{ju\theta}
\end{equation}
with the integer $u=\floor*{-\frac{U-1}{2}},...,\floor*{\frac{U-1}{2}}$, as they fulfill the orthonormality requirement. The notation $[\cdot]_i$ refers to the $i$-th element of a vector, $[\cdot]_{i,j}$ refers to the element in row $i$ and column $j$ of a matrix. From theory \cite{Doron1} it is known that the magnitude of the elements of $H$ decays superexponentially for increasing $u$ beyond $|u|=kr$, where $k$ is the angular wavenumber and $r$ the radius of the smallest sphere enclosing the antenna. Therefore a finite number of coefficients $U$ is sufficient to allow an accurate representation of the antenna pattern. In (\ref{eq:ms}) spatial sampling is defined as a linear operation. The sampling matrix $\bm{H}$ can thus be found by least squares
\begin{equation}
	\hat{\bm{H}} = \bm{E}(\bm{\vartheta}) \bm{\Psi}(\bm{\vartheta})^{\mathrm{H}} \left( \bm{\Psi}(\bm{\vartheta}) \bm{\Psi}(\bm{\vartheta})^{\mathrm{H}} \right)^{-1}.
\end{equation}
Once $\bm{H}$ is found, the interpolation can be performed by (\ref{eq:ms}). For $P$ signals, arriving from angles $\theta_1,...,\theta_P$, the antenna response matrix can be composed as
\begin{equation}
	\bm{A}(\bm{\theta}) = [\bm{a}(\theta_1),...,\bm{a}(\theta_P)].
\end{equation}

\subsection{Antenna Characteristics}\label{ac}
\begin{figure}
  \centering
  \includegraphics[width=\columnwidth,height=0.45\textheight,keepaspectratio]{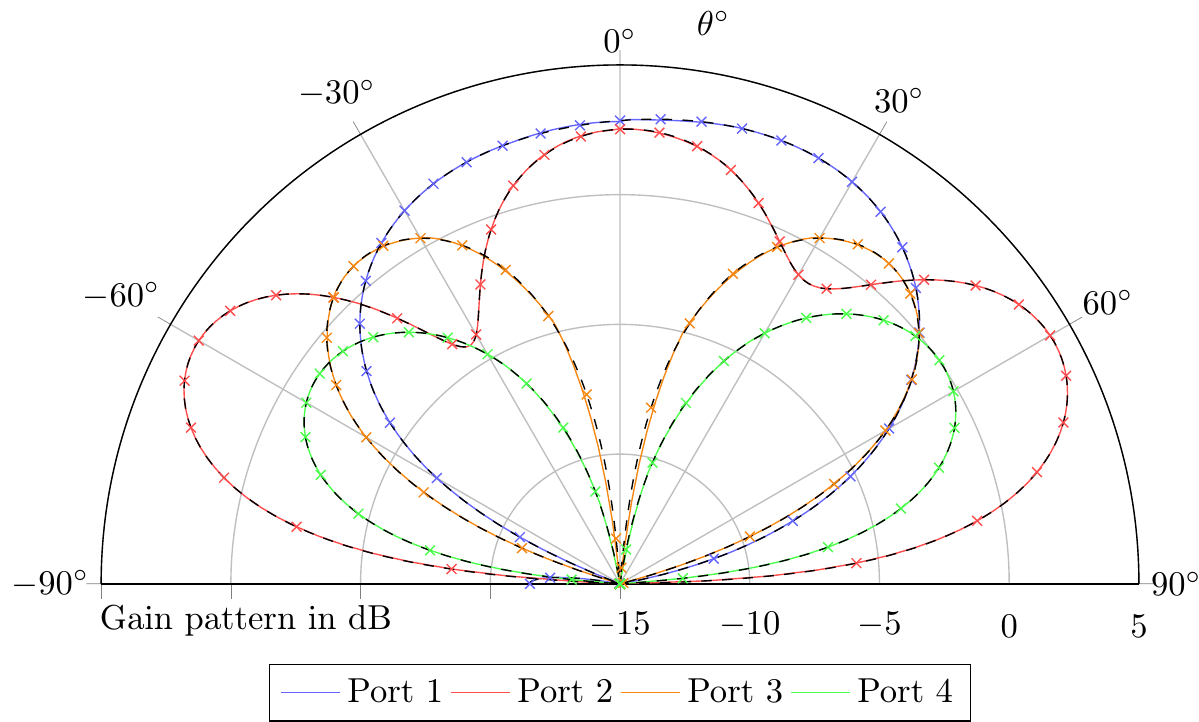}
  \caption{Gain patterns of the investigated MM antenna and the interpolated patterns exploiting both proposed models in the $xz$-plane. The crosses represent the sampled gain patterns of the MM
  antenna prototype. Solid lines represent the gain patterns of the AIT-based model. Dashed lines represent the gain patterns of the WM-based model.}
  \label{fig3}
\end{figure}

The $xz$-plane cut of the gain patterns of the investigated MM antenna prototype as well as the interpolated gain patterns according to the proposed 
models are plotted in Fig.~\ref{fig3}. The simulated EMF samples at angular $5^\circ$ steps are plotted as crosses. 
The interpolated gain patterns of the AIT-based model are plotted as solid lines while the gain patterns of the WM-based model are plotted as dashed lines. 
For the AIT-based model a virtual ULA of $N=4$ elements was positioned on the $z$ axis with an interelement spacing of
$\lambda/4$, where $\lambda$ is the operating wavelength. The mapping coefficients were calculated for a $30^\circ$ sector size. 
When we divide the FoV into sectors we introduce an overlap between adjacent sectors, i.e. neighboring sectors share regions of angular 
samples. This improves the mapping process and allows the choice of larger sector sizes. The overlap size taken in Fig.~\ref{fig3} is 
$15^\circ$. Section~\ref{sec3} discusses the chosen virtual ULA parameters as well as the sector and overlap sizes, in addition 
to their impact on the designed model. The WM-based model was designed using Fourier functions \eqref{eq:Psitheta}. $U=13$ was taken to 
achieve precise interpolation, since the EMF simulation data used for the analysis is quasi noise free. $U$ could be reduced to match the noise floor, 
in case the data would have been obtained from anechoic chamber calibrations. Both models can interpolate the MM antenna simulation data 
very well.

\section{Practical Aspects}\label{sec3}
In this section, we study the effect of the virtual ULA design on the quality of the AIT-based model, as well as the effect of the number of coefficients on the quality of the WM-based model. 
It is known, generally, that errors introduced by various mapping techniques lead to errors in DoA estimations \cite{Belloni07}. 
Accordingly, it is important to carefully design the corresponding model to obtain an accurate mapping of the MM antenna. 

\subsection{Practical Aspects of the AIT-based Model}\label{paAIT}
The design of the virtual array by AIT mapping has been an interesting research topic since its introduction by Bronez \cite{Bronez88}. 
Questions like the optimal number of virtual antenna elements, interelement spacing, and orientation of the virtual ULA are a subject of interest. 
Friedlander in \cite{Friedlander90} chose the virtual array as a ``rule of thumb'' to have elements that are close to the elements of the real array, and a virtual aperture approximately equal to the aperture of the real array. 
B\"uhren \textit{et al.} \cite{Buehren04} studied the relation between virtual array geometry and mapping errors. 
Beside the design of the virtual array, the sector size for the sector-wise interpolation is a crucial parameter. 
On the one hand, large sector sizes are desirable for minimum computational effort. 
On the other hand, the interpolation performs poorly for large sector sizes. 
As suggested in \cite{Friedlander90}, a sector size of $30^\circ$ is commonly used. 

In the following, we analyze the influence of the number of virtual elements, their arrangement (orientation and interelement spacing), 
and the size of sectors and overlaps of the virtual ULA. For the analysis, we take two criteria for assessing the quality of the designed model
into account.
The first criterion is the average transformation error given by
\begin{equation}\label{eq12}
 \xi(\bm{\vartheta}_l)=\frac{{\left\| \bm{G}_l^\mathrm{H} \bm{A}_v(\bm{\vartheta}_l)-\bm{E}(\bm{\vartheta}_l) \right\|}_\mathrm{F}}
 {{\left\|\bm{E}(\bm{\vartheta}_l)\right\|}_\mathrm{F}}.
\end{equation}
Notice that $\xi$ is calculated over all angular samples of the MM antenna at angles $\bm{\vartheta}_l$ within sector $l$.
In the remainder of the paper, we drop the word average, and call $\xi$ the transformation error for simplicity. 
For an accurate model design the transformation error should be kept on the order of $10^{-3}$ or smaller \cite{Friedlander90}. 
The second criterion is to obtain a smooth progression of the interpolated pattern. 
The interpolated gain patterns should not exhibit discontinuities in the sense of rapid variations in a small region, because the original gain patterns illustrated in Fig.~\ref{fig1} do not feature such artifacts.
It is fair to say that for the analysis of the virtual ULA parameters the second criterion holds, but becomes crucial for the analysis of the influence of the FoV sectorization.
We accept the designed model when both criteria are fulfilled.  
A fairly accurate design was found taking the parameters given in Section~\ref{ac}. 
Therefore, in each of the following sections, we alter the parameter being studied in the respective section while keeping the remaining parameters fixed according to the values in Section~\ref{ac}. 

\subsubsection{Influence of the Orientation of the Virtual ULA}\label{oreientation}
Initially, we place the virtual ULA on the $x$ axis where it shares centers with the MM antenna on the center of the Cartesian
coordinate system, as depicted in Fig.~\ref{fig4a}. Next, we place the virtual ULA on the $z$ axis where it again shares centers with the MM antenna on the center of the Cartesian
coordinate system, as depicted in Fig.~\ref{fig4b}.
The array steering vector at arbitrary angle $\theta$ in the first case can be written as
\begin{equation}
\bm{a}_{v,x}(\theta)=[e^{\mathrm{j}kx_1\sin{\theta}},\dots,e^{\mathrm{j}kx_N\sin{\theta}}]^\mathrm{T}.
\end{equation}
\begin{figure}[t]
\includegraphics[width=\columnwidth,height=0.45\textheight,keepaspectratio]{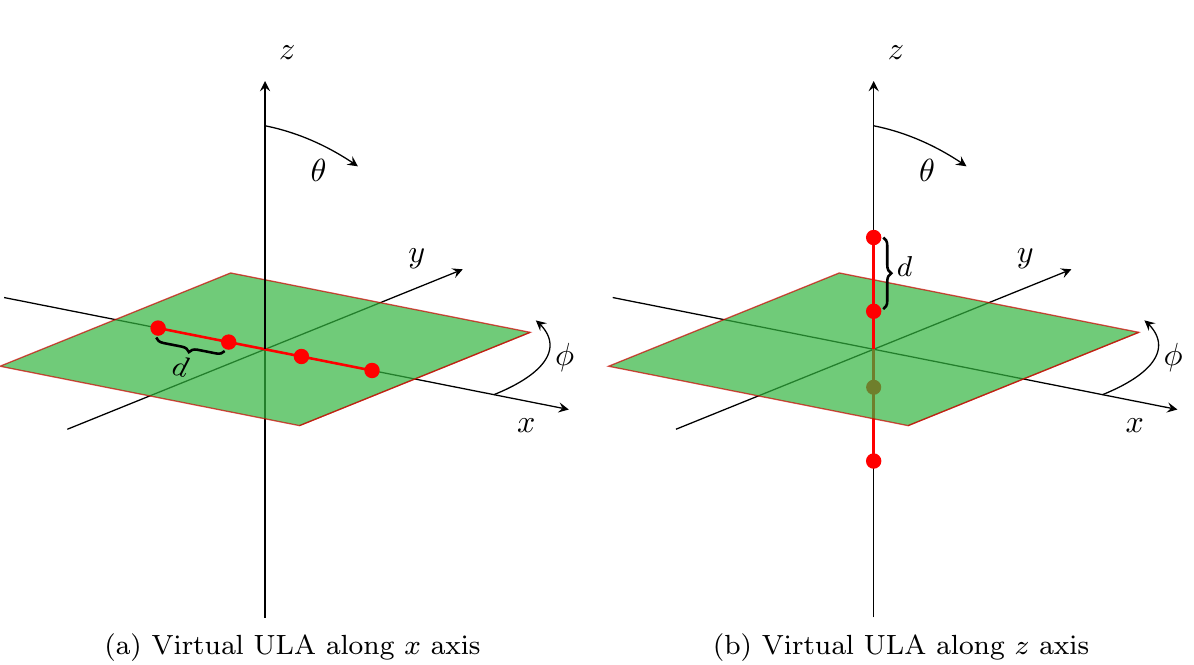}
   \caption{Positioning of the virtual ULA $(a)$ on the $x$ axis and (b) on the $z$ axis.}
   \label{fig4}
\end{figure}
After using this array steering vector to construct the array steering matrix and to perform the mapping, we get the gain patterns plotted in 
Fig.~\ref{fig5}. The figure shows that the model has difficulties matching the sampled pattern of the MM antenna for 
$|\theta|>75^\circ$ (highlighted area), i.e. in the end-fire direction of the virtual ULA. This particularly influences the 
interpolation of mode patterns $2$ and $4$ whose angular samples are not modeled well in the mentioned region. 

Similarly, when the virtual ULA is positioned along the $z$ axis, see Fig.~\ref{fig4b}, the array steering vector at arbitrary angle $\theta$ can be written as
\begin{equation}
\bm{a}_{v,z}(\theta)=[e^{\mathrm{j}kz_1\cos{\theta}},\dots,e^{\mathrm{j}kz_N\cos{\theta}}]^\mathrm{T}.
\end{equation}
The resulting gain patterns are shown in Fig.~\ref{fig5.1}. 
Like the case when the virtual ULA was positioned on the $x$ axis, the model faces difficulties in the end-fire direction of the virtual ULA,
i.e. for $|\theta|<15^\circ$ (highlighted area). 
However, Fig.~\ref{fig5.1} indicates that the interpolation of modes $1$ and $2$ performs well, unlike mode $3$ where the interpolation does not fit the corresponding angular samples well in the mentioned region. 
This can be explained by the behavior of the considered modes. 
First and second modes in the region $|\theta|<15^\circ$ exhibit only a slight change in the progression
of the pattern so the interpolation does not face problems to fit the angular samples. On the other hand, mode $3$, in the same region,
is dropping with a sharp slope towards null. 
Hence, the model performs better when the virtual array is positioned on the $z$ axis rather 
than on the $x$ axis. These results are also confirmed by the transformation error \eqref{eq12}. Taking the mean of the transformation error 
over all sectors yields $\xi_{\mathrm{mean},x}=9.5 \cdot 10^{-3}$ for the virtual ULA positioned on the $x$ axis, and  
$\xi_{\mathrm{mean},z}=1.6 \cdot 10^{-3}$ for the virtual ULA positioned on the $z$ axis, see Fig.~\ref{fig6}. The mean errors differ by almost one order of magnitude.  Therefore, the model 
with the virtual array on the $z$ axis should be preferred.  However, when the precision of the model in the broadside direction is more relevant than
the precision in the end-fire direction, the other orientation may be chosen.
It is generally possible to align the virtual ULA between the $x$ axis and the $z$ axis.
But from our experience, it is advantageous if the virtual ULA is either set on the $x$ axis or the $z$ axis, respectively. 
Otherwise, we risk that the second criterion is violated.
Hence, orientations between the $x$ and the $z$ axis are discarded.

\subsubsection{Influence of the Interelement Spacing}\label{spacing}
It is known from AIT literature that the virtual array elements are preferably placed as close
as possible to the real array elements \cite{Friedlander90}, \cite{Buehren04}. In the case of an MM antenna, the planar radiator itself represents all antenna elements. 
Hence, smaller mapping errors can be expected when the virtual elements are close to the MM antenna structure. 
Fig.~\ref{fig6} shows the transformation error versus the interelement spacing varying between $d=0.1\lambda$ and 
$d=0.5\lambda$ for both orientations. The transformation error is almost constant for the interelement spacing under consideration. Note that $d<0.5\lambda$ does not cause any mutual coupling because $d$ refers to a virtual array, i.e., a mathematical model. Taking into account the size of the 
investigated prototype to be $0.725\lambda \times 0.725\lambda$ at $f_c=7.25$ GHz, it becomes clear that a change of the spacing from  
$d=0.1\lambda$ to $d=0.5\lambda$ is relatively small compared to the area of the MM antenna. 
An interelement spacing $d>0.5\lambda$ was not considered in order to avoid spatial aliasing \cite{Johnson93}.

\begin{figure}
  \centering
\includegraphics[width=\columnwidth,height=0.45\textheight,keepaspectratio]{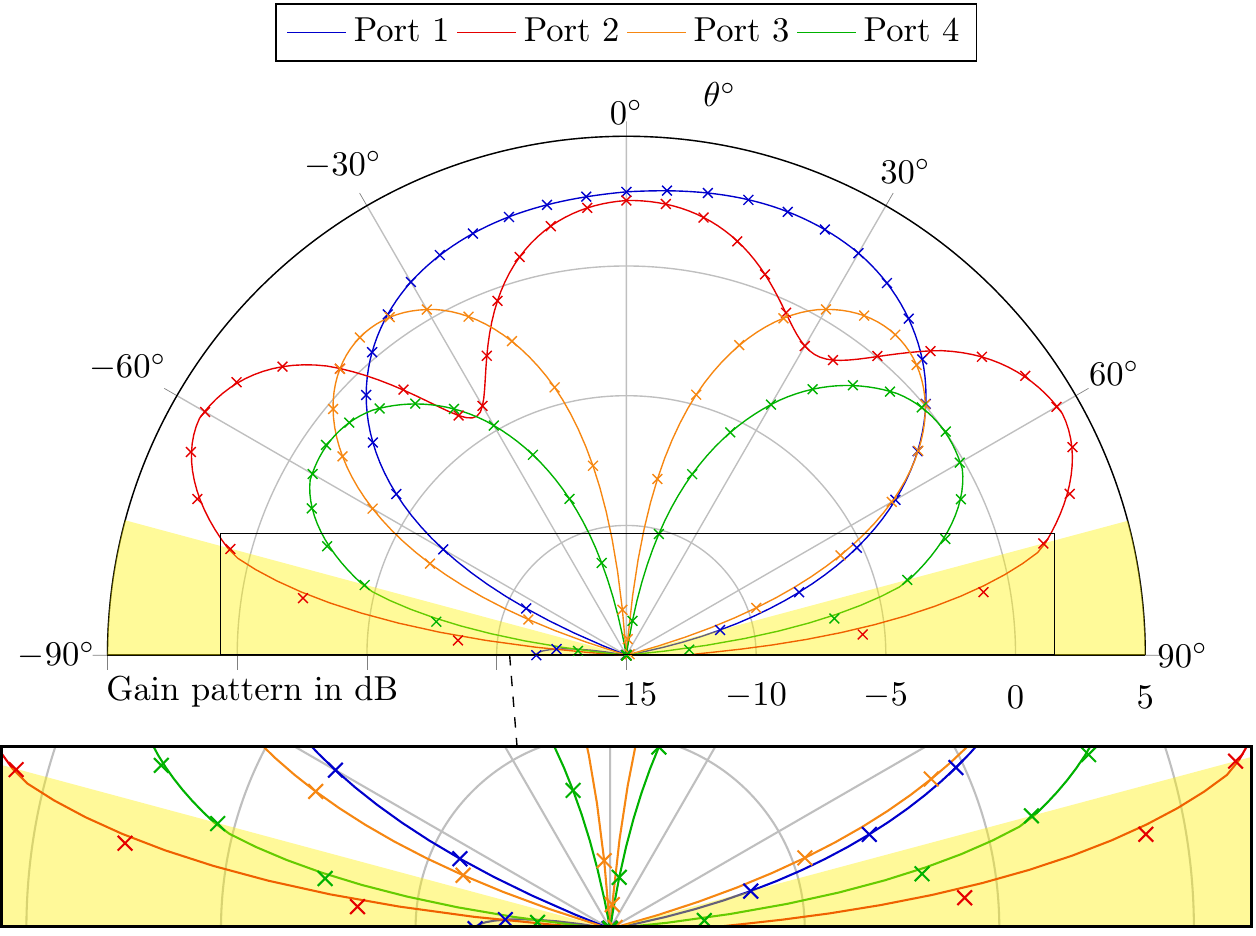}
  \caption{Gain patterns of the investigated MM antenna (crosses) and the interpolated patterns (solid lines) in $xz$-plane. 
   The virtual ULA is positioned on the $x$ axis. The bottom figure is a zoom of the highlighted area of the top part.}
  \label{fig5}
\end{figure}

\begin{figure}
  \centering
\includegraphics[width=\columnwidth,height=0.45\textheight,keepaspectratio]{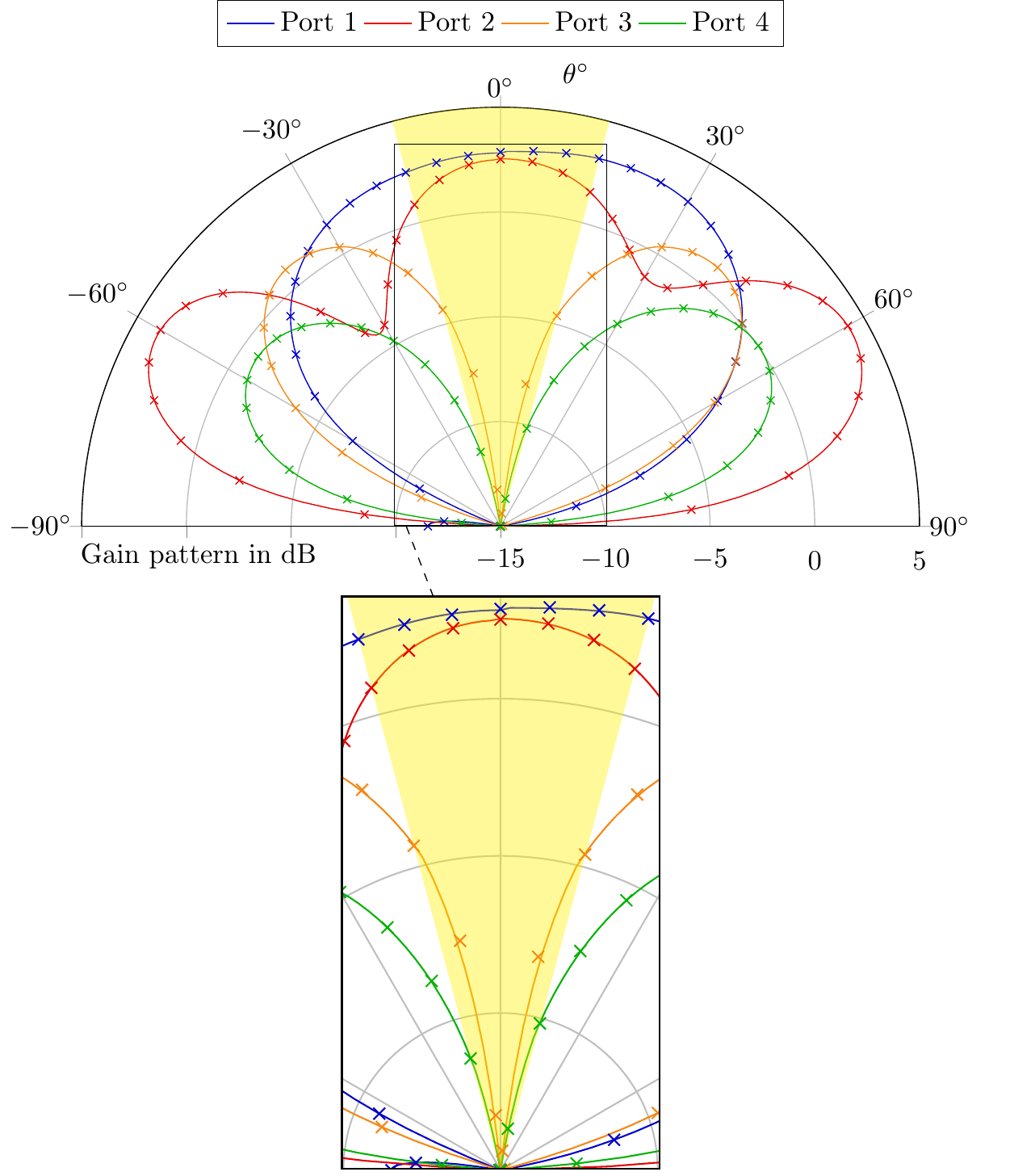}
  \caption{Gain patterns of the investigated MM antenna (crosses) and the interpolated patterns (solid lines) in the $xz$-plane. 
   The virtual ULA is positioned on the $z$ axis. The bottom figure is a zoom of the highlighted area of the top part.}
  \label{fig5.1}
\end{figure}

  \begin{figure}[ht]
     \centering
    \includegraphics[width=\columnwidth,height=0.45\textheight,keepaspectratio]{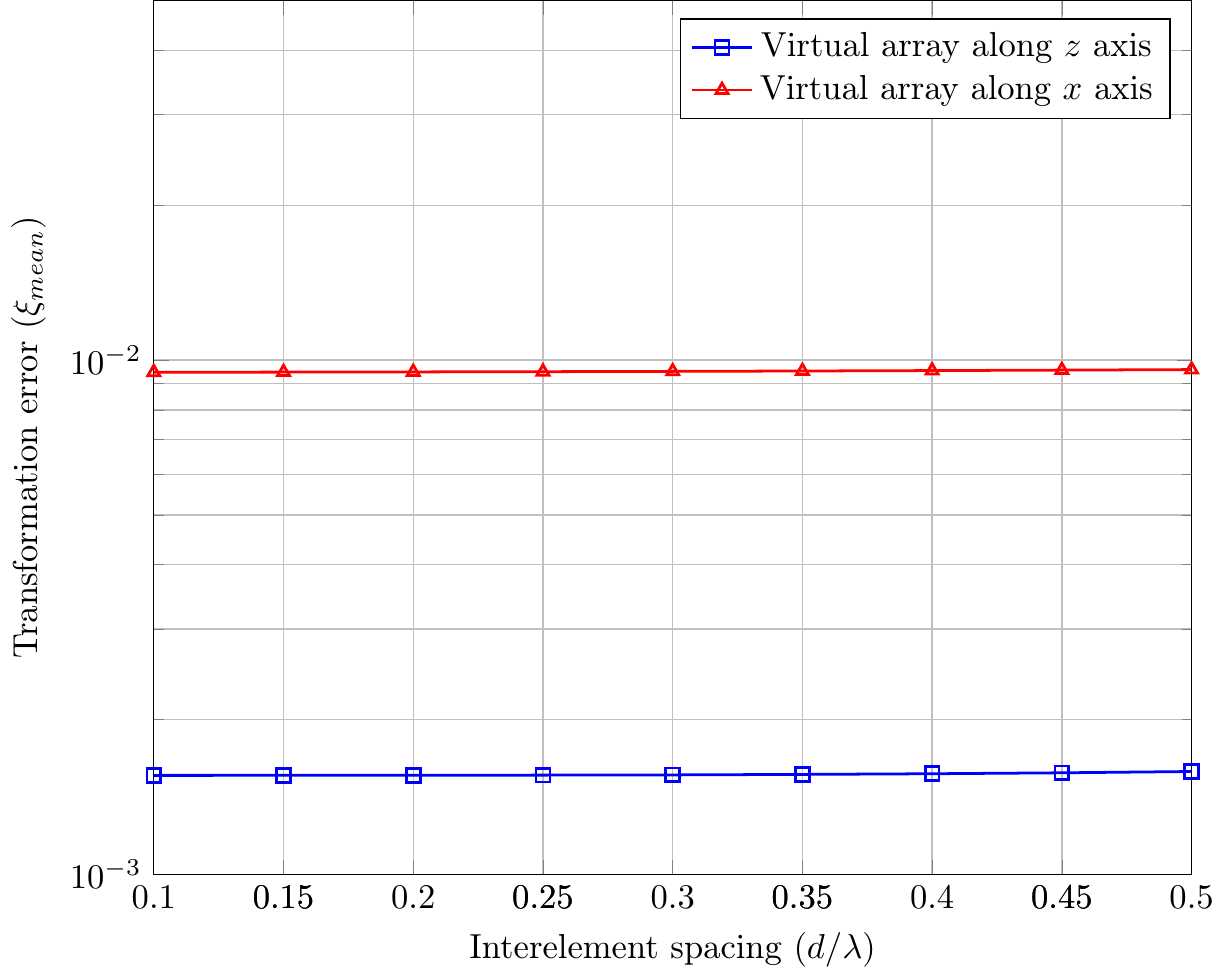}
   \caption{Transformation error versus interelement spacing.}
   \label{fig6}
  \end{figure}

\subsubsection{Influence of the Number of Antenna Elements}
\begin{figure}[ht]
 \centering
\includegraphics[width=\columnwidth,height=0.45\textheight,keepaspectratio]{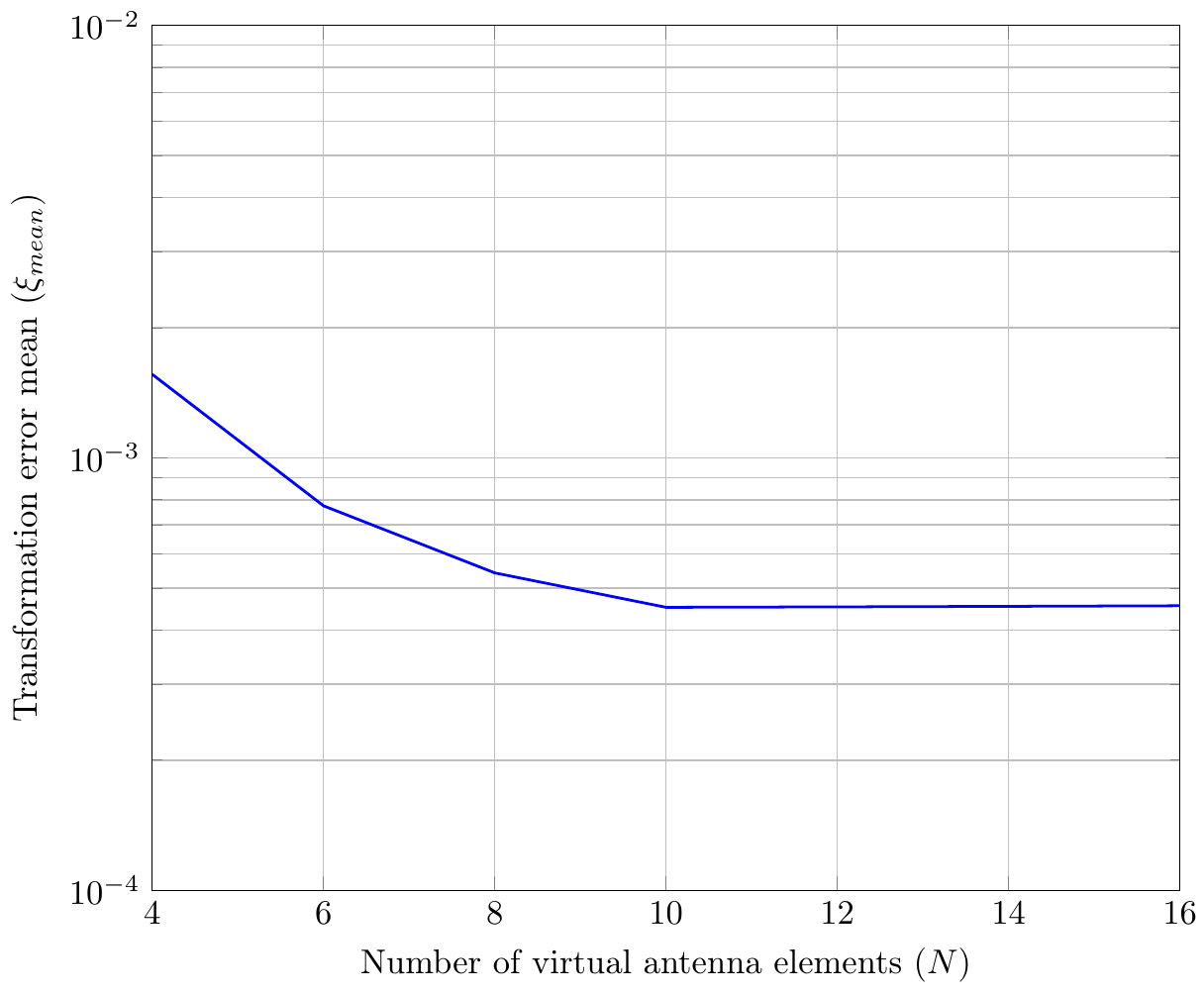}
\caption{Transformation error versus number of virtual antenna elements.}
\label{fig7}
\end{figure}

In order to investigate the impact of the chosen number of virtual antenna elements, the transformation error is plotted versus an increasing number of
virtual elements in Fig.~\ref{fig7}. Starting with the setting shown in Fig.~\ref{fig4b}, in each step two new virtual elements along 
the $z$ axis are added and the transformation error is calculated.
The error curve implies that a better interpolation accuracy is obtained with increasing number of elements. This is expected since a 
larger number of elements results in more degrees of freedom for the mapping matrix. However, after a certain number of virtual elements (about $N=10$),
taking more elements into account does not further reduce the transformation error. This is due to the fact that additional elements move away 
from the structure of the real MM antenna and do not significantly contribute to the interpolation. As a result, the mapping coefficients associated
with these elements have much smaller values compared to the coefficients of the virtual elements closer to the MM antenna.  

As can be seen from \eqref{eq8}, each row of the mapping coefficient matrix contains the mapping coefficients that correspond to a virtual element, while each column contains the mapping coefficients that correspond to a mode of the MM antenna. 
This leads, in case of interpolation with a large number of virtual elements, to rows with very small coefficients in the mapping matrix.
As a result, the mapping matrix could become ill-conditioned, which is likely to cause a large bias of the DoA estimates \cite{Marinho18}. 
Fig.~\ref{fig7} proves that the designed model with $N=M=4$ virtual elements already achieves a quite good transformation.   
  
\subsubsection{Influence of the Sector and Overlap Size}\label{sectorSize}
In the classical AIT algorithm \cite{Friedlander90}, the interpolation sector size is chosen heuristically. 
If the designed model achieves an error in the accepted order, the current design is kept. 
Otherwise, a smaller sector size is taken until an acceptable error is obtained. 
A sector size of $30^\circ$ is common in the literature to achieve an accurate transformation. 
However, studies concerning interpolation techniques using larger sector sizes can be found in \cite{Buehren03} and \cite{Hyberg05}.
With the mentioned sector size, we obtain an error that fulfills the first criterion, but violates the second. 
The designed models have difficulties to obtain a smooth interpolation between the sparsely sampled patterns of the MM antenna. 
For a sector size of $5^\circ$, i.e. for 36 sectors, the model achieves good results.

To enable larger sectors (and hence a lower computational complexity) while keeping an acceptable transformation error and a smooth progression of the interpolated pattern, we propose the concept of overlapping sectors. 
For example, consider a sector $s_1=[0^\circ,30^\circ]$ and the neighboring sector $s_2=[15^\circ,45^\circ]$. 
Then, the overlap region is $s_{ov}=[15^\circ,30^\circ]$. 
Given this example, the entire FoV can be divided into just 11 sectors rather than 36 sectors as the classical AIT algorithm suggests.
After calculating the mapping coefficients for $s_1$ and $s_2$ according to \eqref{eq8}, we end up with two coefficient matrices $\bm{G}_1$ and $\bm{G}_2$ for the overlap region $s_{ov}$.
The optimal coefficient matrix for the overlap region, $\bm{\hat{G}}_{ov}$, is chosen such that the interpolation error is minimized according to
\begin{equation}\label{eq15}
  \bm{\hat{G}}_{ov} = \arg\min_{\left\lbrace\tilde{\bm{G}}_{ov}\in\mathcal{G}\right\rbrace}\left\lbrace \sum_{P_{ov}}
  (\bm{\tilde{G}}_{ov}^\mathrm{H} \bm{a}_v(\vartheta_{p_{ov}})-\bm{\varepsilon}(\vartheta_{p_{ov}}))^2\right\rbrace\  ,
\end{equation}
where the columns of $\bm{\tilde{G}}_{ov}$ are taken independently either from $\bm{G}_1$ or $\bm{G}_2$.
Furthermore, $\vartheta_{p_{ov}}$ is an angular sample in the overlap region and $P_{ov}$ is the number of angular samples in this region.
Remember that $\bm{a}_v(\vartheta_{p_{ov}})$ and $\bm{\varepsilon}(\vartheta_{p_{ov}})$ are the virtual ULA array steering vector and electric field response of the MM antenna at angle $\vartheta_{p_{ov}}$, respectively. 
Notice that no new coefficients are calculated for the overlap regions; rather the best fitting coefficients are chosen from neighboring sectors.
\begin{figure}
  \centering
  \includegraphics[width=\columnwidth,height=0.45\textheight,keepaspectratio]{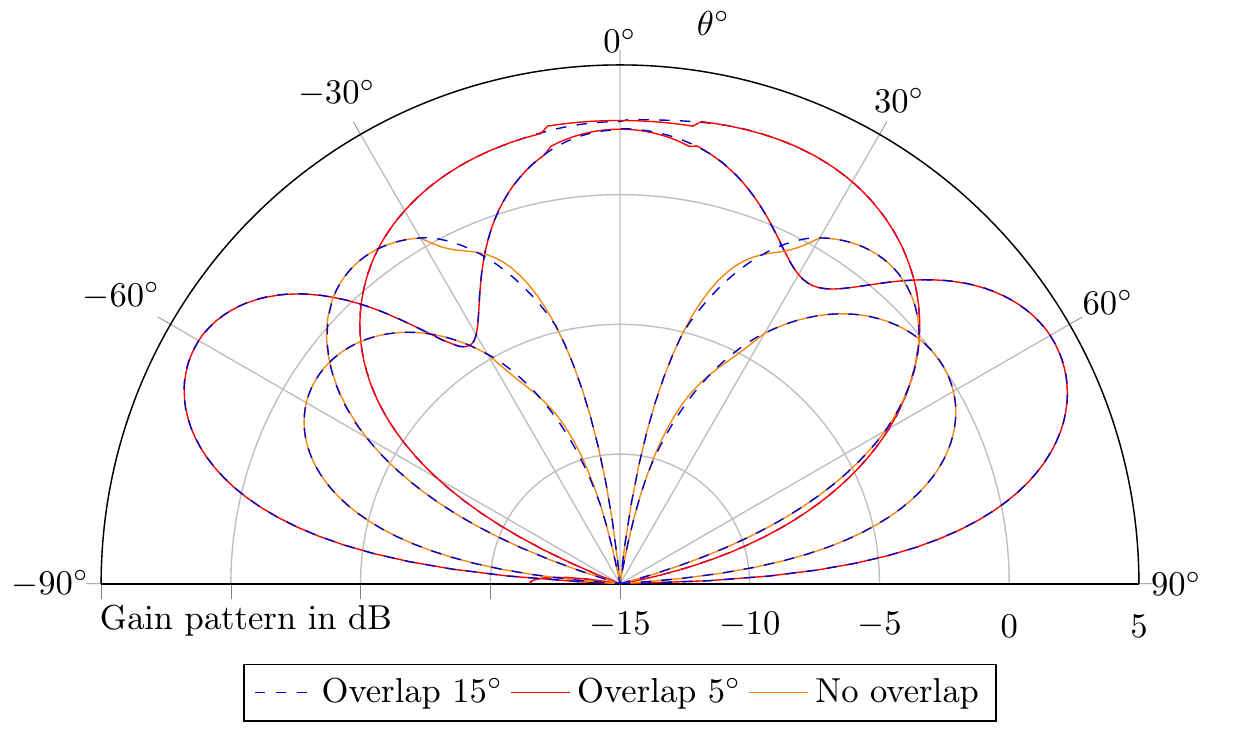}
  \caption{Interpolated gain pattern for different overlap sizes in the $xz$-plane.
          The dashed lines represent the interpolation with $15^\circ$ overlap size. 
          The solid lines of modes $1$ and $2$ represent the interpolation of those modes with $5^\circ$ overlap size, 
          while the solid lines of modes $3$ and $4$ represent the interpolation of those modes without overlaps.}
  \label{fig8}
\end{figure}

A design with $30^\circ$ sector size and $15^\circ$ overlap size (i.e. yielding 11 sectors) obtains a fairly good model accuracy; see Fig.~\ref{fig3}. 
To visualize the impact of an insufficient overlap size, Fig.~{\ref{fig8}} includes interpolated gain patterns for different modes using smaller overlap.
Dashed lines depict the interpolation with $15^\circ$ for comparison. 
Solid lines illustrate an overlap of $5^\circ$ for modes $1$ and $2$ and no overlap for modes $3$ and $4$, respectively.
As indicated by the figure, for $5^\circ$ overlap the model is not accurate in the region $|\theta|<15^\circ$ for modes $1$ and $2$. 
The same applies for modeling without overlap in the region $15^\circ<|\theta|<30^\circ$ for modes $3$ and $4$.

\subsection{Practical Aspects of the WM-based Model}
\subsubsection{Influence of the Number of Coefficients}\label{ss:coef}
\begin{figure}[t]
	\centering
	\includegraphics{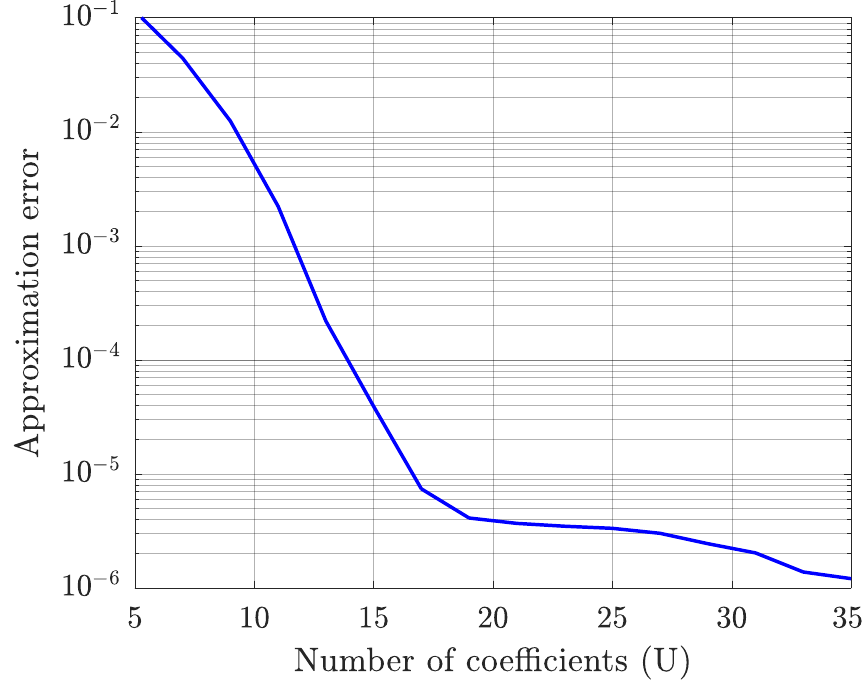}
	\caption{Approximation error versus number of coefficients $U$ for wavefield modeling. Beyond $U = 26$ the coefficient magnitude decays superexponentially.}
	\label{fig:approxError_wm}
\end{figure}

For the WM approach introduced in Section~\ref{swm}, a critical design parameter is the number of Fourier coefficients, i.e. basis functions, $U$. The approximation error of the antenna characteristic is defined analogously to (\ref{eq12}) as
\begin{equation}\label{eqr1}
	\xi(\bm{\vartheta})=\frac{{\left\| \bm{H} \, \bm{\Psi}(\bm{\vartheta}) - \bm{E}(\bm{\vartheta}) \right\|}_\mathrm{F}}
	{{\left\|\bm{E}(\bm{\vartheta})\right\|}_\mathrm{F}}.
\end{equation}
Fig.~\ref{fig:approxError_wm} shows the approximation error versus the number of Fourier coefficients. It can be seen that by increasing the number of coefficients, the approximation error becomes smaller. For $U<19$ the error drops fast, whereas for $19 \geq U \geq 35$ the increase in approximation accuracy is significantly smaller.

\section{Model-based DoA Estimation}\label{sec4}
In this section, we apply maximum-likelihood-based DoA estimation to both antenna models and study the influence of the choice of parameters on the DoA root mean squared error (RMSE). 
First, we start with the considered signal model. 
Assuming that $Q<M$ narrowband signals \cite{Stoica88} arriving from angles $\bm{\theta}=[\theta_1,...,\theta_q]$, one snapshot of the output signal $\bm{y}\in\mathbb{C}^{N\times1}$ can be expressed as
\begin{equation}\label{sigmod}
 \bm{y}(k) = \bm{A}(\bm{\theta})\bm{x}(k)+\bm{n}(k),\qquad k = 1,\dots,K, 
\end{equation}
where $K$ is the number of snapshots, $\bm{A}(\bm{\theta})\in\mathbb{C}^{M\times Q}$ is the antenna response, $\bm{x}(k)\in\mathbb{C}^{Q\times 1}$ is the signal vector arriving from angles $\bm{\theta}$, and $\bm{n}(k)\in\mathbb{C}^{M\times1}$ is a complex valued zero mean Gaussian distributed white noise process with variance $\sigma^2$ and covariance matrix $\sigma^2\mathbf{I}$. 
The array steering vectors in $\bm{A}(\bm{\theta})$ are assumed to be linearly independent \cite{Wax85}, and the number of sources $Q$ is assumed to be known \cite{Ziskind88}.
The case of an unknown number of sources has been studied in \cite{Wax85Kai}. 
Given the observations $\bm{y}(k)$, we estimate the angles ${\bm{\theta}=[\theta_1,...,\theta_q]}$ of the $Q$ sources, based on the maximum-likelihood estimator presented in \cite{Wax85}. 

The joint probability density function of the $K$ observations can be written as
\begin{multline}
  p(\bm{y}(1),\dots,\bm{y}(K)) = \prod^K_{k=1}\frac{1}{\pi \det [\sigma^2\mathbf{I}]} \\
  \cdot \exp \left(-\frac{1}{\sigma^2} \left\| \bm{y}(k)- \bm{A}(\bm{\theta})\bm{x}(k)\right\|^2\right),
\end{multline}
where $\mathrm{det}[\cdot]$ denotes the determinant.
After neglecting the constant terms, the log-likelihood function becomes  
\begin{equation}
 L = K \log \sigma ^2 - \frac{1}{\sigma^2} \sum^K_{k=1} \left\|\bm{y}(k)- \bm{A}(\bm{\theta})\bm{x}(k)\right\|^2.
\end{equation}
Next, the maximization of the log-likelihood function with respect to the unknown parameters $\bm{\theta}$ and $\bm{x}(k)$ leads to the following multi-variate minimization problem:
\begin{equation}
[\hat{\bm{\theta}},\hat{\bm{x}}(k)]= \arg\min_{\tilde{\bm{\theta}},\tilde{\bm{x}}(k)} \left\lbrace\sum^K_{k=1} \left\|\bm{y}(k)- \bm{A}(\tilde{\bm{\theta}})\tilde{\bm{x}}(k)\right\|^2\right\rbrace.
 \label{minill}
\end{equation}
Assuming (for the time being) the angles $\bm{\theta}$ to be known, the least squares solution with respect to $\bm{x}(1),\dots,\bm{x}(K)$ can be expressed by
\begin{equation}
\label{LS_signal}
 \hat{\bm{x}}(k) =  \left( \bm{A}(\bm{\theta})\bm{A}^\mathrm{H}(\bm{\theta}) \right)^{-1}\bm{A}(\bm{\theta})^\mathrm{H} \bm{y}(k).
\end{equation}
Substituting (\ref{LS_signal}) into (\ref{minill}) yields the following minimization problem:
\begin{equation}\label{eq16}
 \hat{\bm{\theta}}=\arg\min_{\tilde{\bm{\theta}}} \left\lbrace\sum^K_{k=1} \left\|\bm{y}(k)- \bm{\Pi}_{\bm{A}(\tilde{\bm{\theta}})}\bm{y}(k)\right\|^2\right\rbrace,
\end{equation}
where $\bm{\Pi}_{\bm{A}(\bm{\theta})}$ is the projection matrix onto the space spanned by the vectors of $\bm{A}(\bm{\theta})$:
\begin{equation}
 \bm{\Pi}_{\bm{A}(\bm{\theta})} = \bm{A}(\bm{\theta})\bm{A}^\dagger(\bm{\theta})
  = \bm{A}(\bm{\theta})\left( \bm{A}(\bm{\theta})\bm{A}^\mathrm{H}(\bm{\theta}) \right)^{-1}\bm{A}(\bm{\theta})^\mathrm{H}.
\end{equation}
The projection matrix of the orthogonal complement of the column space of $\bm{A}(\bm{\theta})$ is by definition
\begin{equation}\label{eq18}
 \bm{\Pi}^\perp_{\bm{A}(\bm{\theta})} \overset{\Delta}{=} \mathbf{I} -  \bm{\Pi}_{\bm{A}(\bm{\theta})}.
\end{equation}
Therefore, after substituting (\ref{eq18}) into (\ref{eq16}) and exploiting the trace operator, $\mathrm{tr}(\cdot)$, the angle estimates can be rewritten as
\begin{equation}\label{eq20}
\hat{\bm{\theta}}=\arg\min_{\tilde{\bm{\theta}}} \sum^K_{k=1} \mathrm{tr}[\bm{\Pi}^\perp_{\bm{A}(\tilde{\bm{\theta}})}\bm{y}(k)\bm{y}(k)^\mathrm{H}].
\end{equation}
Finally, by defining the sample covariance matrix as
\begin{equation}
 \hat{\bm{R}} = \frac{1}{K}\sum_{k=1}^K\bm{y}(k)\bm{y}(k)^\mathrm{H},
\end{equation}
the minimization problem (\ref{eq20}) can be formulated as
\begin{equation}
 \hat{\bm{\theta}}=\arg\min_{\tilde{\bm{\theta}}} \sum^K_{k=1} \mathrm{tr}[\bm{\Pi}^\perp_{\bm{A}(\tilde{\bm{\theta}})}\hat{\bm{R}}].
\end{equation}

After having defined the signal model and the considered maximum-likelihood estimator, we perform numerical simulations to verify the influence of different parameters of both models on RMSE of the DoA estimates. 
For providing a fair comparison, in the following simulations the received signal $\bm{y}(k)$ in \eqref{sigmod} was generated using the quantized EMF data, while the estimator in \eqref{minill} employs either the AIT-based model, see Section~\ref{sec2.1}, or the WM-based model, see Sec~\ref{swm}. 
Each of the numerical simulations is performed with $K=1000$ snapshots and 
\begin{equation}
 \text{RMSE}(\theta)=\sqrt{\frac{1}{N_{\mathrm{MCr}}}\sum_{n_{\mathrm{MCr}}=1}^{N_{\mathrm{MCr}}}(\hat{\theta}_\mathrm{MCr}-\theta)^2},
\end{equation}
where $N_{\mathrm{MCr}} = 1000$ Monte Carlo runs have been conducted at $\text{SNR}=20$~dB, unless stated otherwise. 
Afterwards, the mean of the estimated DoA RMSE is calculated over $\theta\in[-90^\circ,90^\circ]$.
The parameters for the following analysis are those given in Section~\ref{ac}.
The parameter under investigation in the respective section will be altered, while the remaining parameters will be fixed.  

\subsection{RMSE versus Number of Antenna Elements}
\begin{figure}[ht]
 \centering
\includegraphics[width=\columnwidth,height=0.45\textheight,keepaspectratio]{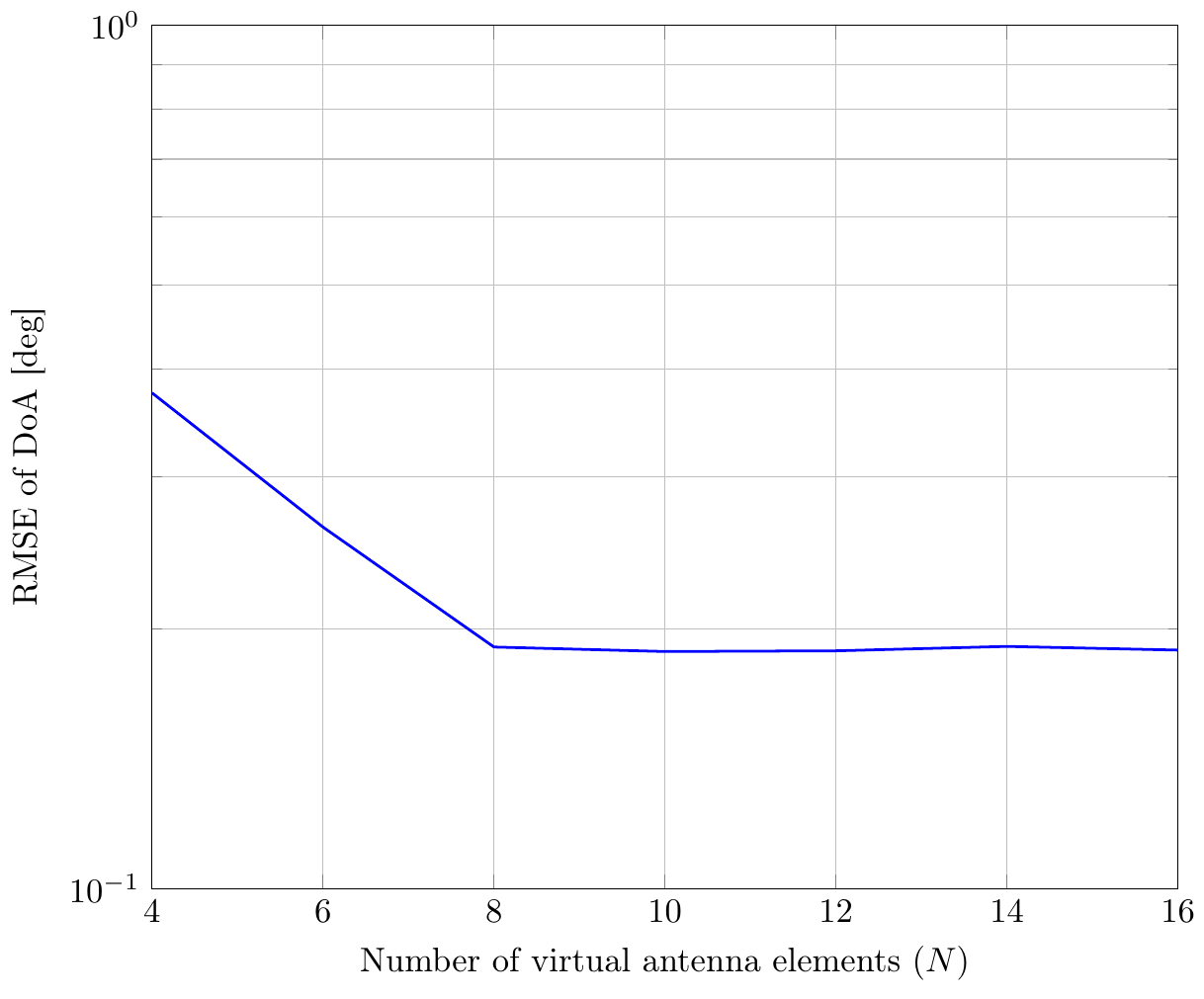}
\caption{RMSE versus number of virtual antenna elements.}
\label{fig11}
\end{figure}

To study the impact of the number of virtual elements on the DoA estimation, the RMSE of the estimates versus the number of virtual elements is shown in Fig.~\ref{fig11}.
As expected, a larger number of virtual elements delivers better DoA estimates since the transformation error decreases, see Fig.~\ref{fig7}.
However, after a certain number of virtual elements (about $N=10$), there will not be any significant improvement.
This result coincides with the results shown in Fig.~\ref{fig7}, where it is clear that more elements do not reduce the transformation error.

\subsection{RMSE versus Orientation of the Virtual ULA}
\begin{figure}[ht]
 \centering
  \includegraphics[width=\columnwidth,height=0.45\textheight,keepaspectratio]{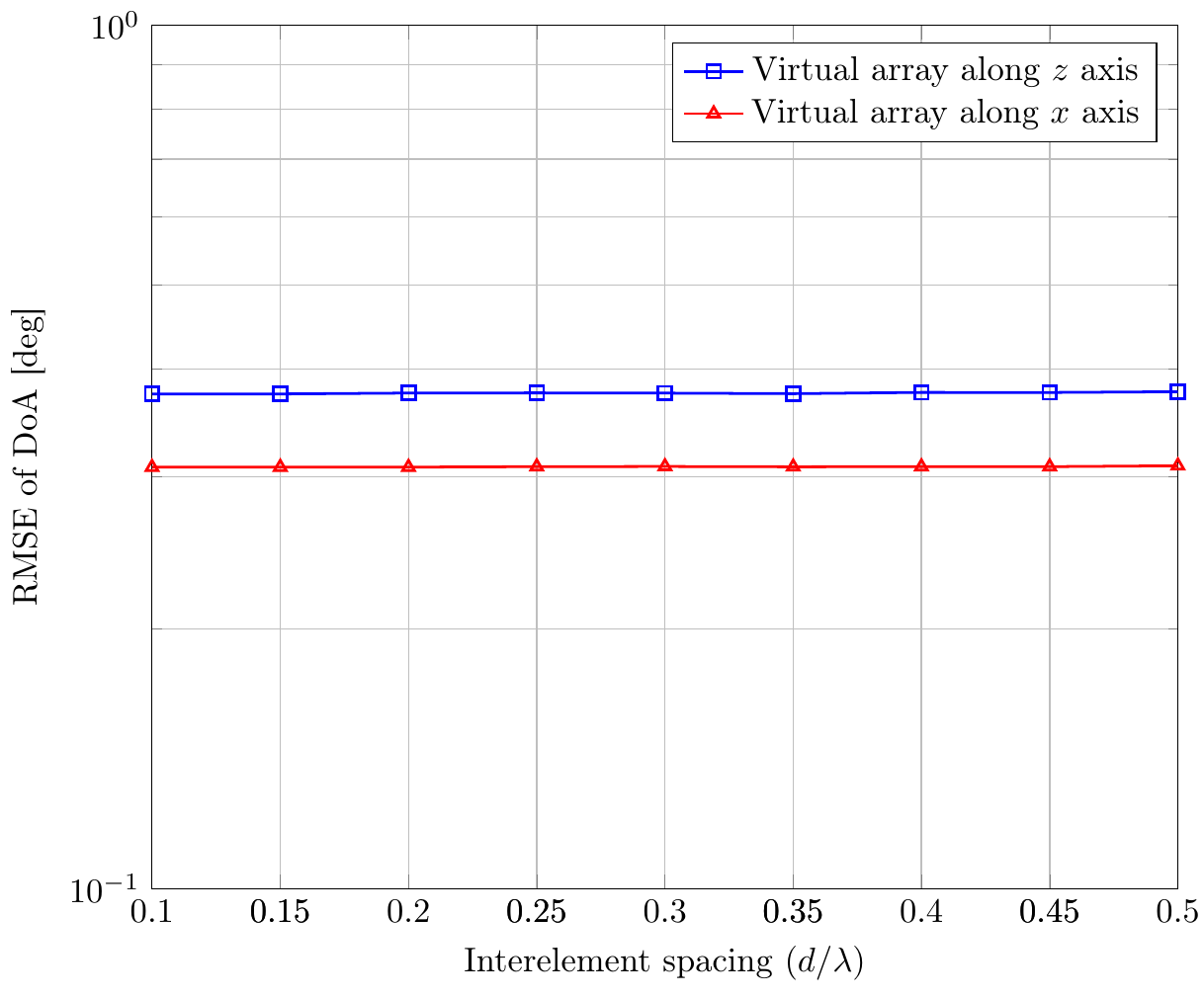}
\caption{RMSE versus interelement spacing of virtual antenna elements.}
\label{fig12}
\end{figure}

Results of numerical simulations for a virtual array placed along the $x$ axis and $z$ axis for different interelement spacing are given in Fig.~\ref{fig12}.
As the interelement spacing between $d=0.1\lambda$ and $d=0.5\lambda$ does not much affect the transformation error, see Section~\ref{spacing}, the RMSE change with respect to interelement spacing is also insignificant.
The interesting observation in Fig.~\ref{fig12} is that the model with the virtual array along the $x$ axis obtains, on average, smaller RMSE than the model with virtual array along the $z$ axis, 
despite the fact that the transformation error in the first case is larger than in the second case. 
The explanation can be found in Fig.~\ref{fig5} and Fig.~\ref{fig5.1}.
In the case of positioning the virtual array on the $x$ axis, the transformation error in the $|\theta|>75^\circ$ region is associated with low gain for all modes. 
In the case of positioning the virtual array on the $z$ axis, the transformation error in the $|\theta|<15^\circ$ region is associated with high gain for modes $1$ and $2$. 
Therefore, the impact of the transformation error on the estimator in the mentioned region becomes larger. 
However, the RMSE difference between both models is in order of $10^{-2}$, which is insignificant for a DoA RMSE in the order of $10^{-1}$. 

\subsection{RMSE versus Sectorization and Overlapping Size}
\begin{figure}[t]
 \centering
\includegraphics[width=\columnwidth,height=0.45\textheight,keepaspectratio]{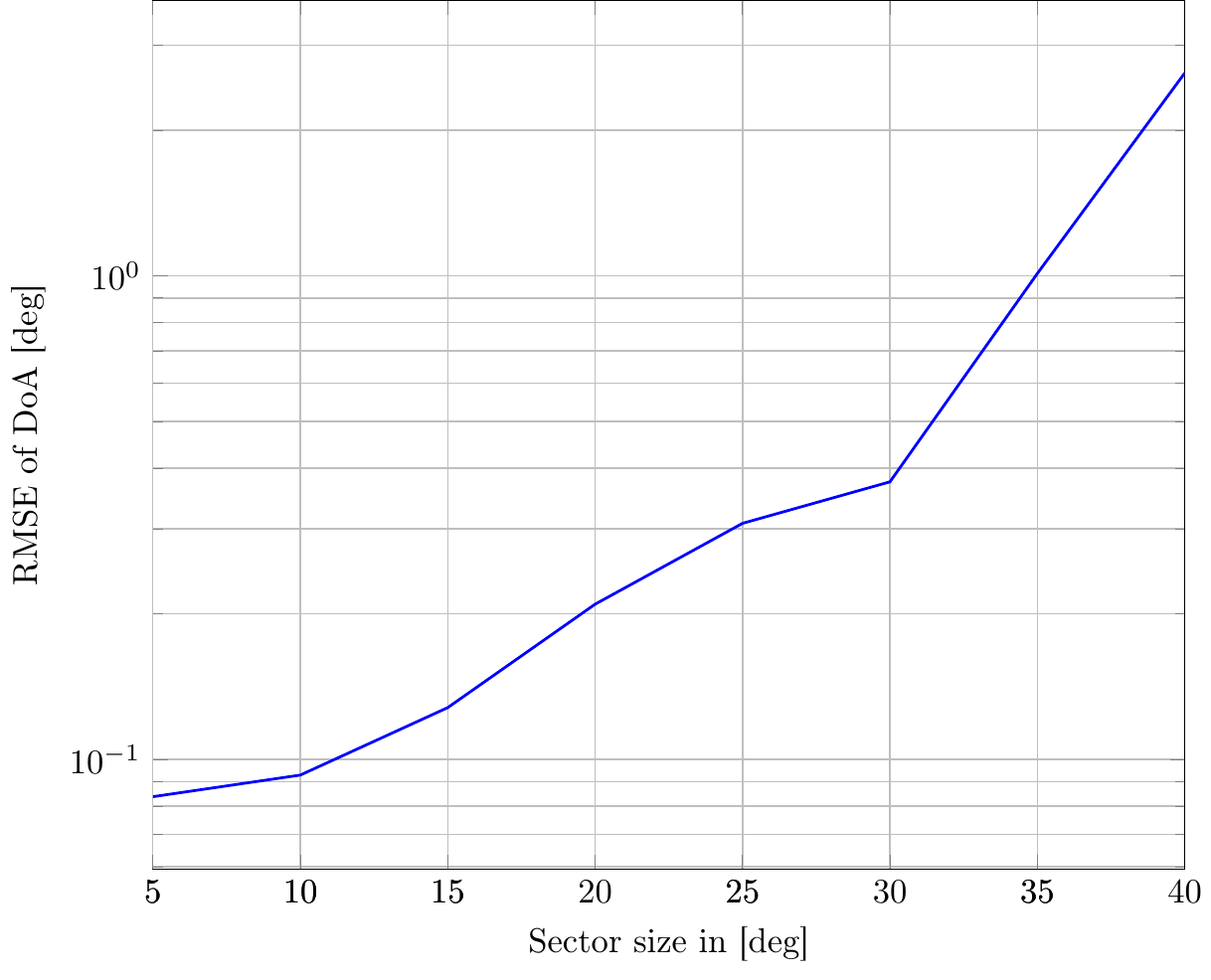}
\caption{RMSE versus sector size.}
\label{fig13}
\end{figure}

The influence of the chosen sector and overlap size is analyzed by considering different sector sizes, each with $50\%$ overlap, except for the sector size of $5^\circ$ where the overlap is $0^\circ$.  
Results are depicted in Fig.~\ref{fig13}. It can be verified from the figure that the DoA RMSE increases with increasing sector size.
\noindent
As the designed model for small sector sizes up to $30^\circ$ fulfills both quality criteria defined in Section~\ref{paAIT}, the estimator performs well.
For larger sector sizes the model violates the quality criteria introducing a rapid increase in DoA RMSE.  

\subsection{RMSE versus Number of Coefficients for Wavefield Modeling}
\begin{figure}[t]
	\centering
	\includegraphics{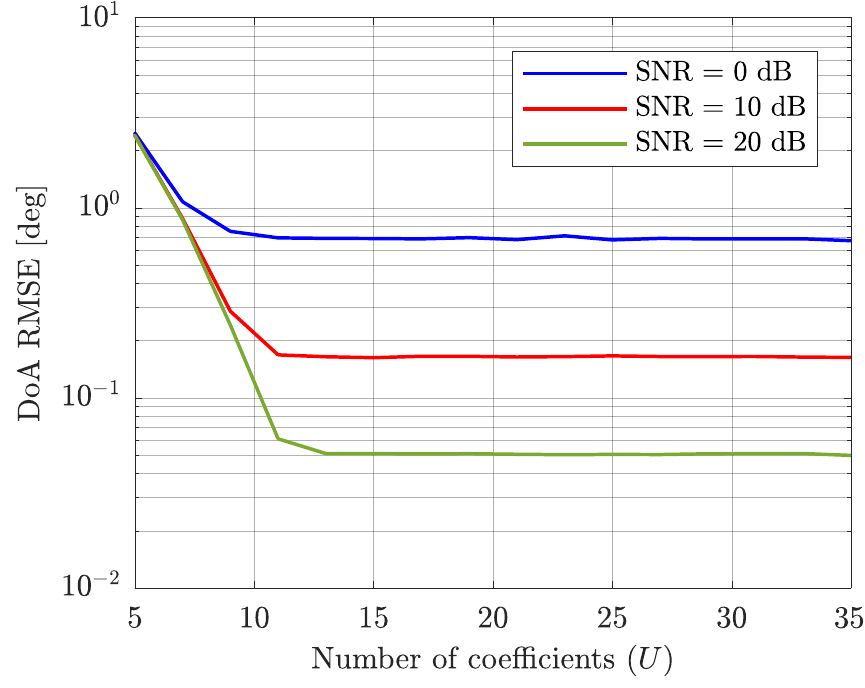}
	\caption{DoA estimation RMSE averaged over $\theta$ for $\mathrm{SNR}=0, 10, 20$~dB versus number of coefficients $U$ for wavefield modeling.}
	\label{fig:doa_rmse_wm}
\end{figure}

With Fig.~\ref{fig:doa_rmse_wm} we want to answer the question of how many Fourier coefficients are necessary in practice for the WM approach introduced in Section \ref{ss:coef}. Again, the received signals are generated based on the original EMF simulation data with $5^\circ$ grid and the estimator employs the WM-based model with a different number of coefficients $U$, see (\ref{eq:Psitheta}). As the Fourier functions in (\ref{eq:Psitheta}) are symmetric around $0$, only odd numbers of coefficients $U$ are used. 
% The estimation RMSE is averaged over $\theta \in [-90^\circ,90^\circ]$ and calculated for $1000$ Monte Carlo runs. 
The plot shows that for $U<13$, a model mismatch causes an increased RMSE. This model mismatch becomes more severe the less coefficients are used. For $U>13$, increasing the number of coefficients does not reduced the RMSE. Apparently with $U=13$ the model is good enough even for $\text{SNR}=20$~dB, such that the estimation accuracy is limited by the noise instead of the model.

\section{Discussion}\label{sec5}
For the purpose of modeling an MM antenna we have studied the array interpolation technique and the waveform modeling technique.
The first method models the MM antenna as a virtual ULA, whereas the second method describes the response of the MM antenna based on a mathematical model.
While both methods are able to model the MM antenna well, each method has pros and cons.
The AIT-based model is more intuitive.  It offers the possibility to apply computationally efficient DoA methods like ESPRIT \cite{Roy89} and Root-MUSIC \cite{Friedlander93}.
Nevertheless, this method suffers from highly correlated signals outside the considered sector, which leads to a degradation of the DoA estimates \cite{lau04}.
The WM-based method can model the MM antenna response over the whole FoV without sectorization, but can only be applied to DoA estimators of higher complexity.

The accuracy of an MM antenna model, either based on AIT or WM, has been shown to depend on the choice of parameters of the respective model.
In the case of AIT, the option of placing a virtual ULA with four virtual elements on the $z$ axis, given a interelement spacing of $\lambda/4$, a sector size of $30^\circ$, and an overlap size of $15^\circ$, provides a good approximation.
For these parameters, a set of eleven mapping coefficient matrices, each of size $4\times4$, is obtained. 
These matrices are delivered in the appendix.  Given these matrices, the 2-D antenna characteristics of all four modes of the MM prototype under investigation can be reproduced by the community in interpolated form with high angular resolution, without having access to EMF data samples. 
In the case of WM, when choosing the Fourier functions in \eqref{eq:Psitheta} as basis functions, $U=13$ Fourier coefficients for each port are sufficient to obtain a good approximation.  The corresponding parameters are delivered in the appendix as well. 

An analysis of DoA performance utilizing a maximum-likelihood DoA estimator has revealed the dependency of the RMSE on the accuracy of the designed antenna model. 
In general, transformation errors are biasing the DoA estimates. 
The bias introduced by the AIT-based model strongly depends on the sectorization. 
A wrong choice of the sector size increases the RMSE severely, while the bias resulting from parameters of the virtual array still obtains acceptable accuracy in terms of the RMSE. 
On the other hand, the accuracy of the WM-based model improves with increasing number of coefficients $U$. 
Beyond $U=13$ coefficients the RMSE does not improve significantly.
The WM-based model achieves smaller RMSE values at high SNR values, because for the AIT-based model it is known that at high SNRs the impact of the transformation error may be larger than that of the noise \cite{Marinho15}.
The studied and discussed parameters of the AIT and WM-based models to obtain minimum RMSE apply to the MM antenna prototype introduced in\mbox{\cite{Manteuffel16}}.
However, the presented guidelines for choosing those parameters are also valid for MM antennas having a different number of modes.
For the AIT-based model, parameters such as number of elements and interelement spacing of the virtual array are less critical than the orientation of the virtual array and the sectorization.
The latter two parameters depend on the radiation patterns of the considered modes.
For the WM-based model, the number of coefficients depends on the electrical size of the antenna \mbox{\cite{Doron1}}. For an MM antenna with a different number of modes but the same electrical size, the same number of coefficients can be used.
Moreover, in analogy to increasing the number of antenna elements of a conventional antenna array, having more modes would improve the performance of the maximum-likelihood estimator of the DoA. 
However, care should be taken in this case to insure a sufficiently small correlation between the modes.
%having a different number of modes of the MM antenna corresponds to having a different number of antenna elements in an array and would affect the DoA estimation. For instance, having more modes improves the performance of the maximum-likelihood estimator.}

\section{Conclusion}\label{sec6}
In the areas of array signal processing and digital communications, a huge amount of results are available for uniformly-spaced antenna arrays.
Although multi-mode antennas are of increasing interest because they mimic antenna arrays, this antenna type is arbitrary and currently not well modeled from a signal processing/communications point of view.
As a solution, we adopt two modeling methods to multi-mode antennas: The array interpolation technique and wavefield modeling.
% The array interpolation technique is able to match the parameters of a virtual uniform linear antenna array to the antenna response of a given multi-mode antenna.
Consequently, for the purpose of signal processing one could replace the multi-mode antenna by the virtual antenna. 
The wavefield modeling method is conceptually similar with a higher degree of abstraction. 
Both models are able to interpolate the given antenna pattern. 
This is an important practical feature because electromagnetic field data is frequently available only in spatially quantized form. 
As a possible application, both modeling concepts are applied to coherent maximum-likelihood direction-of-arrival estimation. 
The impact of parameter sets and modeling errors are studied and compared.
Finally, optimized parameter sets are provided. 

\section{Acknowledgments}

This work has been funded by the German Research Foundation (DFG) under contract numbers HO~2226/17-1 and FI~2176/1-1.  
The authors would like to thank Prof.~D.~Manteuffel from the University of Hannover, Germany, for a collaboration on MM antennas within a related project.  

\bibliographystyle{IEEEtran}
\bibliography{references}

 \newpage
 \appendices
\twocolumn[
\section{AIT mapping coefficients}\label{sec14}
According to the AIT-based model introduced in Section~\ref{sec2.1} and the results in Section~\ref{paAIT}, the investigated $M=4$ port MM antenna is modeled by a virtual ULA of $N=4$ elements along the $z$ axis with interelement spacing $d=\lambda/4$.
A $30^\circ$ sector size with $15^\circ$ overlap size is considered.
This results in $L=11$ sectors.
In the following, we provide the corresponding mapping coefficient matrices $\bm{G}_l \in \mathbb{C}^{N \times M}$:]
% In this appendix we provide the mapping coefficients for the AIT based model introduced in section \ref{sec2.1}. 
% The investigated $M=4$ port MM antenna According to the results in Section~\ref{paAIT}, a virtual ULA of $N=4$ elements along the $z$ axis with interelement spacing $d=\lambda/4$,  $30^\circ$ sectors with $15^\circ$ overlap size were considered. 
% This results in $L=11$ sectors and, correspondingly, mapping coefficients matrices $\bm{G}_l \in \mathbb{C}^{N \times M}$.
% The investigated MM prototype has $M=4$ antenna ports. On the basis of Figures \ref{fig:approxError_wm} and \ref{fig:doa_rmse_wm} we haven chosen $U=13$ coefficients, which is sufficient to yield a reasonable accuracy.
\begin{alignat*}{2}
& \spaceto{\bm{G}_11}{\bm{G}_1} &&=
\begin{pmatrix}
 \begin{array}{rrrr}
-0.3379929-0.02271694i & -1.42895+1.202409i & -0.9751865-0.3748656i & -0.492186+0.5581889i \\
0.4176824-1.154261i & 1.967421-2.31803i & 1.625532+0.1412085i & 0.2301585-0.6351565i \\
0.3223047+2.440937i & -2.122732+2.68573i & -0.336772+0.6634518i & -0.2600317+0.7019622i \\
-0.3296267-1.199872i & 1.669175-1.511143i & -0.3515941-0.5677777i & 0.5343497-0.6326347i 
\end{array}
\end{pmatrix} \\[2pt]
% \end{align*}
% \vspace{-0.3cm}
% \begin{align*}
& \spaceto{\bm{G}_11}{\bm{G}_2} &&=
\begin{pmatrix} 
 \begin{array}{rrrr}
-0.5257688-0.2182654i & -1.491053+0.5900567i & -1.825524+0.00023299i & -0.4804487+0.2481393i \\
1.259226-0.9855009i & 3.160347-0.860535i & 3.039504-2.184621i & 0.7197898+0.1457578i \\
-0.5164995+2.803788i & -3.980786+2.111863i & -0.1668957+3.351573i & -1.107282+0.3307682i \\
-0.1562431-1.477924i & 2.319373-1.702403i & -0.9127693-1.272015i & 0.8426109-0.6902792i  
\end{array}
\end{pmatrix} \\[2pt]
% \end{align*}
% \vspace{-0.3cm}
% \begin{align*}
& \spaceto{\bm{G}_11}{\bm{G}_3} &&=
\begin{pmatrix}
 \begin{array}{rrrr}
0.2355258-0.966885i & -0.07919153+1.1492i & -3.819897+0.2823089i & -0.2614507-0.1482683i \\
1.751606+2.164831i & -0.6700898+1.642975i & 5.824912-7.471549i & 1.281117+1.35567i \\
-3.378592+1.361043i & -3.812815-2.525889i & 2.48541+8.6982i & -2.404057+0.07166003i \\
0.8047454-1.980577i & 3.579044-0.7417844i & -2.892205-1.601538i & 1.167957-0.9885241i 
\end{array}
\end{pmatrix} \\[2pt]
% \end{align*}
% \vspace{-0.3cm}
% \begin{align*}
& \spaceto{\bm{G}_11}{\bm{G}_4} &&=
\begin{pmatrix}
 \begin{array}{rrrr}
2.229434-1.579864i & 1.498453+4.051515i & -10.61997-2.782043i & -0.125851-2.746992i \\
1.461651+8.369543i & -10.40638+3.099267i & 21.04782-23.75545i & 8.472205+4.167886i \\
-9.11039-1.005892i & -1.850187-12.16477i & 13.01172+28.40181i & -7.314702+5.981984i \\
2.199137-3.518214i & 6.388087+0.9812878i & -10.30044-0.3892399i & -0.1883448-3.165292i 
\end{array}
\end{pmatrix} \\[2pt]
% \end{align*}
% \vspace{-0.3cm}
% \begin{align*}
& \spaceto{\bm{G}_11}{\bm{G}_5} &&=
\begin{pmatrix}
 \begin{array}{rrrr}
22.54393-6.21377i & 21.10923+5.911715i & -149.361-150.026i & 21.91906-90.69182i \\
7.309108+70.42268i & -23.36855+60.32424i & 511.5459-379.0759i & 261.0971+103.3414i \\
-71.35295-3.269566i & -56.67663-32.23264i & 302.445+560.5287i & -138.0371+243.8143i \\
5.6482-24.02199i & 15.33167-16.24202i & -199.0336+72.6838i & -73.38952-56.65479i 
\end{array}
\end{pmatrix} \\[2pt]
% \end{align*}
% \vspace{-0.3cm}
% \begin{align*}
& \spaceto{\bm{G}_11}{\bm{G}_6} &&=
\begin{pmatrix}
 \begin{array}{rrrr}
2.566821-2.167523i & -2.216303+11.71209i & 0.272016-0.064465i & -0.1733718+0.06848533i \\
3.905379+8.974951i & -31.26493-10.59543i & 0.176097+0.694365i & 0.01236176-0.4497193i \\
-9.801546+2.145884i & 13.70103-30.85554i & -0.684591+0.155507i & 0.3838582+0.1305691i \\
0.8692102-3.941048i & 11.85216+6.540964i & -0.0337077-0.252317i & -0.04299996+0.09484828i 
\end{array}
\end{pmatrix} \\[2pt]
% \end{align*}
% \vspace{-0.3cm}
% \begin{align*}
& \spaceto{\bm{G}_11}{\bm{G}_7} &&=
\begin{pmatrix}
 \begin{array}{rrrr}
-17.58397+4.597868i & -20.80397+11.14517i & 149.5724+150.1123i & -22.21876+90.8226i \\
-7.569878-53.19476i & -20.59792-66.47453i & -511.8159+379.4558i & -261.0655-104.1034i \\
52.66765-0.4211724i & 68.2275-11.48275i & -302.7949-560.8167i & 138.6718-243.5577i \\
-1.280433+16.5098i & 2.572562+23.77483i & 199.158-72.84539i & 73.30874+56.80175i 
\end{array}
\end{pmatrix} \\[2pt]
% \end{align*}
% \vspace{-0.3cm}
% \begin{align*}
& \spaceto{\bm{G}_11}{\bm{G}_8} &&=
\begin{pmatrix}
 \begin{array}{rrrr}
0.4377447-0.4678576i & -0.1257663+5.391872i & 10.7628+2.832154i & -0.07823439+2.886467i \\
0.6393967+2.282524i & -11.53063-2.535386i & -21.17842+23.95025i & -8.511159-4.650072i \\
-3.448172+0.6891519i & 3.32417-10.81164i & -13.20107-28.52284i & 7.682717-5.839174i \\
0.7302143-2.147879i & 5.186405+2.139559i & 10.36128+0.2717766i & 0.1592258+3.229746i
\end{array}
\end{pmatrix} \\[2pt]
% \end{align*}
% \vspace{-0.3cm}
% \begin{align*}
& \spaceto{\bm{G}_11}{\bm{G}_9} &&=
\begin{pmatrix}
 \begin{array}{rrrr}
-0.2823835-0.5094339i & -0.6978094+1.757147i & 3.876744-0.2328124i & 0.1008805+0.2457422i \\
1.553832+0.2893502i & -0.679947-0.4472327i & -5.865568+7.425067i & -1.246506-1.672012i \\
-1.892054+2.295319i & -2.359054-1.248481i & -2.47812-8.651438i & 2.589976+0.08344666i \\
0.02546755-1.850211i & 2.800566-0.7596384i & 2.877324+1.526914i & -1.179903+0.9938523i
\end{array}
\end{pmatrix} \\[2pt]
% \end{align*}
% \vspace{-0.3cm}
% \begin{align*}
& \spaceto{\bm{G}_11}{\bm{G}_{10}} &&=
\begin{pmatrix}
 \begin{array}{rrrr}
-0.7278701+0.01933685i & -1.896334+0.9618478i & 1.857428+0.07028304i & 0.3429652-0.193996i \\
1.039633-1.720876i & 3.33959-2.033801i & -3.090063+2.040726i & -0.6228015-0.331919i \\
0.06188963+3.052901i & -3.370072+3.00317i & 0.2611086-3.257506i & 1.152816-0.20234i \\
-0.5702037-1.45945i & 1.810973-1.870706i & 0.867896+1.212707i & -0.8196431+0.663218i 
\end{array}
\end{pmatrix} \\[2pt]
% \end{align*}
% \vspace{-0.3cm}
% \begin{align*}
& \bm{G}_{11} &&=
\begin{pmatrix}
 \begin{array}{rrrr}
-0.4583179+0.0411382i & -1.703869+1.378376i & 1.019873+0.4343056i & 0.3747402-0.5523541i \\
0.2939612-1.326862i & 2.152771-2.79889i & -1.690817-0.2366195i & -0.1032964+0.6001909i \\
0.4985535+2.281285i & -1.903679+3.010688i & 0.41739-0.6188696i & 0.1973064-0.6822424i \\
-0.5523052-1.145579i & 1.376495-1.595895i & 0.3197086+0.520304i & -0.460818+0.6157851i 
\end{array}
\end{pmatrix} \\
\end{alignat*}

\clearpage
 \twocolumn[
\section{WM sampling matrix}\label{sec15}
In this appendix we provide the sampling matrix $\bm{H} \in \mathbb{C}^{M \times U}$ for the wavefield modeling approach introduced in Section~\ref{swm}, considering the same MM prototype with $M=4$ antenna ports. On the basis of Figures~\ref{fig:approxError_wm} and \ref{fig:doa_rmse_wm} we haven chosen $U=13$ coefficients, which is sufficient to yield a reasonable accuracy.]
\begin{alignat*}{2}
 &  \spaceto{[\bm{H}]_{1:4,9:12}}{[\bm{H}]_{1:4,1:4}} &&=
  \begin{pmatrix}
  \begin{array}{rrrrr}
    0.01235-0.001501i & 0.06108+0.04358i & -0.2633-0.02637i & 0.4341+0.2486i \\ 
    0.01408-0.1033i & 0.06935+0.3663i & -0.06287-1.067i & -0.167+2.661i  \\ 
    0.01478-0.05369i & -0.08709+0.1286i & 0.1088-0.3239i & -0.06128+0.8119i \\ 
    0.0007279+0.01641i & -0.01672-0.03058i & -0.0261-0.02027i & -0.04259-0.086i \\
    \end{array}
  \end{pmatrix} \\[7pt]
  & \spaceto{[\bm{H}]_{1:4,9:12}}{[\bm{H}]_{1:4,5:8}} &&=
  \begin{pmatrix}
    \begin{array}{rrrrr}
    -0.8519-0.8543i & 1.041+0.7495i & -1.235-1.646i & 0.9176+0.744i   \\ 
    -0.4496-3.444i & 0.3933+4.271i & -0.7591-5.002i & 0.3498+4.325i  \\ 
    0.185-0.504i & -0.04566+0.6791i & 0.0514-0.01475i & -0.03447-0.6254i  \\ 
    0.3964+0.3361i & 0.02951-0.08018i & -0.07423-0.01255i & 0.1301+0.07376i   \\ 
    \end{array}
  \end{pmatrix} \\[7pt]
  &[\bm{H}]_{1:4,9:12} &&=
  \begin{pmatrix}
    \begin{array}{rrrrr}
    -0.8669-0.928i & 0.4244+0.4084i & -0.2688-0.08001i & 0.06771+0.06359i  \\ 
    -0.478-3.481i & -0.1735+2.79i & -0.07445-1.094i & 0.07408+0.36i  \\ 
    -0.1257+0.4581i  & -0.002854-0.7975i & -0.08404+0.3119i & 0.07907-0.1239i  \\ 
    -0.5114-0.3655i & 0.1024+0.1278i & -0.005089+0.01658i & 0.02246+0.02706i  \\ 
    \end{array}
  \end{pmatrix} \\[7pt]
  &  \spaceto{[\bm{H}]_{1:4,9:12}}{[\bm{H}]_{1:4,13}} &&=
  \begin{pmatrix}
    \begin{array}{rrrrr}
    0.01242-0.01148i \\ 
    0.01391-0.1094i \\ 
    -0.01208+0.05338i \\ 
    -0.001222-0.01641i \\ 
    \end{array}
  \end{pmatrix}
\end{alignat*}

\end{document}